
\input phyzzx
\catcode`\@=11 
\def\NEWrefmark#1{\step@ver{{\;#1}}}
\catcode`\@=12 
%

\def\square{\kern1pt\vbox{\hrule height 1.2pt\hbox{\vrule width 1.2pt\hskip 3pt
   \vbox{\vskip 6pt}\hskip 3pt\vrule width 0.6pt}\hrule height 0.6pt}\kern1pt}

\def\bra#1{\langle #1 |}
\def\ket#1{| #1 \rangle}
\def\vev#1{\langle #1 \rangle}
\def\for{{\rm for}}

\def\ov{{\overline}}
\def\bU{{\bf U}}
\def\A{{\cal A}}
\def\B{{\cal B}}
\def\C{{\cal C}}
\def\D{{\cal D}}
\def\H{\widehat{\cal H}}
\def\HH{{\cal H}}

\def\K{\wh{\cal K}}

\def\M{{\cal M}}

\def\O{{\cal O}}
\def\P{{\cal P}}

\def\s{{\cal S}}

\def\V{{\cal V}}

\def\p{\partial}

\def\ov{\overline}

\def\wt{\widetilde}
\def\wh{\widehat}

\def\B{{\cal B}}

\def\P{{\cal P}}
\def\V{{\cal V}}
\def\O{{\cal O}}
\def\s{{\cal S}}
\def\p{\partial}
\overfullrule=0pt
\baselineskip 13pt plus 1pt minus 1pt
\nopubblock
{}~ \hfill \vbox{\hbox{MIT-CTP-2244, TIFR-TH-93-37}
\hbox{IP/BBSR/93-56, November 1993}
\hbox{hep-th/9311009} }\break
\titlepage
\title{QUANTUM BACKGROUND INDEPENDENCE OF}
\titlestyle{CLOSED STRING FIELD THEORY}
\author{Ashoke Sen \foot{E-mail  address: sen@tifrvax.tifr.res.in,
sen@tifrvax.bitnet } \foot{Permanent address: Tata Institute of
Fundamental Research,
Homi Bhabha Road, Bombay 400005, India. } }
\address{Institute of Physics, Sachivalaya Marg, Bhubaneswar 751005, India}
\andauthor
{Barton Zwiebach \foot{E-mail address: zwiebach@irene.mit.edu,
zwiebach@mitlns.bitnet.\hfill\break Supported in part by D.O.E.
contract DE-AC02-76ER03069.}}
\address{Center for Theoretical Physics,
LNS and Department of Physics\break
MIT, Cambridge, Massachusetts 02139, U.S.A.}

\abstract{ We prove local background independence of the complete
quantum closed string field theory using the
recursion relations for string vertices and the existence of
connections in CFT theory space.
Indeed, with this data we construct an antibracket preserving
map between the state spaces of two nearby
conformal theories taking the corresponding
string field measures  $d\mu e^{2S/\hbar}$ into each other.
A geometrical  construction of the map is achieved by introducing
a Batalin-Vilkovisky (BV) algebra on spaces of Riemann surfaces,
together with a map to the BV algebra of string functionals.
The conditions of background independence show
that the field independent terms of the master action
arise from vacuum vertices $\V_{g,0}$, and that the overall
$\hbar$-independent normalization of the string field
measure involves the theory space connection.
Our result puts on firm ground the widely believed statement that
string theories built from nearby conformal theories are different states
of the same theory.}

\endpage

\def\define#1#2\par{\def#1{\Ref#1{#2}\edef#1{\noexpand\refmark{#1}}}}
\def\con#1#2\noc{\let\?=\Ref\let\<=\refmark\let\Ref=\REFS
         \let\refmark=\undefined#1\let\Ref=\REFSCON#2
         \let\Ref=\?\let\refmark=\<\refsend}

\let\refmark=\NEWrefmark

\define\zwiebachlong{B. Zwiebach, `Closed string field fheory: Quantum
action and the Batalin-Vilkovisky master equation', Nucl. Phys {\bf B390}
(1993) 33, hep-th/9206084.}

\define\ghoshalsen{D. Ghoshal and A. Sen, `Partition function of
perturbed minimal models and background dependent free energy of string
theory', Phys. Lett. {\bf B265} (1991) 295.}

\define\spivak{M. Spivak, ``Calculus on Manifolds'', Benjamin/Cummings
publishing (1965).}

\define\kugosuehiro{T. Kugo and K. Suehiro,  `Nonpolynomial closed
string field theory: action and gauge invariance', Nucl.
Phys. {\bf B337} (1990) 434.}

\define\polchinski{J. Polchinski, `One loop string
path integral', Comm. Math. Phys. {\bf 104} (1986) 37.}

\define\senmatrix{A. Sen, `Matrix models and gauge invariant field theory of
subcritical strings', Int. J. Mod. Phys. {\bf A7} (1992) 2559.}

\define\seibergshenker{N. Seiberg and S. Shenker, `A note on background
(in-)dependence, Phys. Rev. {\bf D45} (1992) 4581, hepth/9201017.}

\define\sonodazwiebach{H. Sonoda and B. Zwiebach, `Closed string field theory
loops with symmetric factorizable quadratic differentials',
Nucl. Phys. {\bf B331} (1990) 592.}

\define\senone{A. Sen, `On the background independence of string
field theory', Nucl. Phys. {\bf B345} (1990) 551}

\define\sentwo{A. Sen,`On the background independence of
string field theory (II). Analysis of on-shell $S$-matrix elements',
Nucl. Phys. {\bf B347} (1990) 270}

\define\senthree{A. Sen, `On the background independence of string field
theory (III). Explicit field redefinitions',
Nucl. Phys. {\bf B391} (1993) 550, hep-th/9201041}

\define\hatazwiebach{H. Hata and B. Zwiebach, `Developing the covariant
Batalin-Vilkovisky approach to string theory', MIT-CTP-2184, to appear in
Annals of Physics. hep-th/9301097.}

\define\wittenqbi{E. Witten, `Quantum background independence in
string theory',
IASSNS-HEP-93/29, June 1993, hep-th/ 9306122.}

\define\ooguri{ M. Bershadsky, S. Cecotti, H. Ooguri, and C. Vafa,
``Holomorphic anomalies in topological field theories,'' HUTP-93/A008,
hep-th/9302103.}

\define\oogurii{ M. Bershadsky, S. Cecotti, H. Ooguri, and C. Vafa,
``Kodaira-Spencer theory of Gravity, and Topological Field theory,''
HUTP-93/A008, hep-th/9309140.}

\define\kugozwiebach{T. Kugo and B. Zwiebach, `Target space duality
as a symmetry of string field theory',  Prog. Theor. Phys. {\bf 87}
(1992) 801, hep-th/9201040.}

\define\rangaconnection{K. Ranganathan, `Nearby CFT's in the operator
formalism: The role of a connection', to appear in Nucl. Phys. B.
hep-th/9210090.}

\define\rangasonodazw{K. Ranganathan, H. Sonoda and B. Zwiebach,
`Connections on the state-space over conformal field theories',
MIT-preprint MIT-CTP-2193, April 1993, hep-th/9304053.}

\define\thornpr{C. B. Thorn, `String field theory', Phys. Rep.
{\bf 174} (1989) 1.}

\define\schwarz{A. Schwarz, `Geometry of Batalin-Vilkovisky quantization', UC
Davis preprint, hep-th/9205088, July 1992. }

\define\alvarez{L. Alvarez-Gaume, C. Gomez, G. Moore and C. Vafa,
`Strings in the operator formalism', Nucl.
Phys. {\bf B303} (1988) 455;\hfill\break
C. Vafa, `Operator formulation on Riemann surfaces',
Phys. Lett. {\bf B190} (1987) 47.}

\define\senzwiebach{A. Sen and B. Zwiebach, `Local background independence
of classical closed string field theory', MIT preprint, CTP\#2222,
to appear in Nucl. Phys. B,  hep-th/9307088.}

\define\senzwiebachgauge{A. Sen and B. Zwiebach, `A note on gauge
transformations in Batalin-Vilkovisky theory', MIT preprint, CTP\#2240,
hep-th/9309027.}

\define\zwiebachoc{B. Zwiebach, `Quantum open string theory with manifest
closed string factorization' Phys. Lett. {\bf 256 B} (1991) 22; `Interpolating
string field theories', Mod. Phys. Lett. {\bf A7} (1992) 1079. }

\define\wittenab{E. Witten, `A note on the antibracket formalism', Mod.
Phys. Lett. {\bf A5} (1990) 487.}

\define\getzler{E. Getzler, `Batalin-Vilkovisky algebras and two-dimensional
topological field theories', MIT Math preprint, hep-th/9212043.}

\define\lianzuckerman{B. H. Lian and G. Zuckerman,
``New Perspectives on the BRST-algebraic structure of string theory'',
hep-th/9211072, Yale preprint, November 1992.}

\define\penkavaschwarz{M.~Penkava and A.~Schwarz,
``On Some Algebraic Structures Arising in String Theory",
UC Davis preprint, UCD-92-03, hep-th/9212072.}

\define\senzwiebachnew{A. Sen and B. Zwiebach, ``A BV algebra on Moduli
spaces of Riemann surfaces'', to appear.}

\define\schwarznew{A. S. Schwarz, ``Symmetry transformations in
Batalin-Vilkovisky formalism'', UC Davis preprint, October 1993,
hep-th/9310124.}

\chapter{Introduction and Summary}

Closed string field theory can be formulated
as a full quantum theory in a completely precise way. Indeed,
the master action
for closed string fields has finite $\hbar$-dependent terms which are
determined by BRST symmetry in a way that is logically independent of
the issue of perturbative divergences (see [\zwiebachlong] ).
This is precisely
in the spirit of Batalin-Vilkovisky (BV) quantization, where the aim is to find
a well defined quantum master action, possibly $\hbar$-dependent, which
determines the quantum theory.
The  situation in string field theory is in
contrast with the situation in particle field theory where, typically,
the issue of perturbative divergences becomes mixed with the problem of finding
a suitable master action.  Given that quantum
closed string theory is precisely defined, and, its formulation requires a
choice of a conformal field theory representing a background, the question of
background independence can be addressed.
In the present paper we define what
we believe is the physically correct notion for quantum background
independence of a theory written in the BV formalism. We then prove
that quantum closed string field theory,
for the case of nearby backgrounds related by marginal deformations,
satisfies the proposed notion of background independence.
This paper should be considered the sequel of Ref.[\senzwiebach] where
we proved local background independence of classical closed string
field theory.
The main tool for the present work is a BV algebra that we define
on subspaces of moduli spaces of Riemann surfaces, together with
a map to the BV algebra of string functionals, respecting the algebraic
structure. Proving background independence establishes that string
theories constructed around nearby conformal theories are, in fact, different
states of the same theory. This widely believed statement has therefore
been put on a firm ground. We hope that the insights gained in proving
background independence will be useful in constructing a manifestly
background independent formulation of string field theory.

The notion of background independence of a classical field theory
is familiar. A classical theory is manifestly background independent if
it is written without having to choose a consistent background representing
a classical solution. In such case no particular
classical solution plays any special role. An example
is provided by classical gravitation.
Einstein's action is written for an arbitrary metric representing an
arbitrary space-time background. No Ricci flat background plays a role
in the formulation of the theory.
Alternatively, the formulation of a
theory may require
choosing a background corresponding to a classical solution. In this case the
theory is not manifestly background independent since, {\it a priori}, the
theory depends on the background that has been used to formulate the theory.
One may prove background independence by
formulating the theory using two different backgrounds, and then showing
that there is a classical solution in one of the theories such that, after
a shift of the field by this classical solution, one obtains, up to a field
redefinition, the theory formulated in the other background.
This was the way background independence of classical closed string field
theory was first discussed [\con\senone\sentwo\senthree\noc].

In Ref.[\senzwiebach]
we used the BV formulation of classical closed string field theory to prove
local background independence. We showed that there is a string field
transformation that takes the classical master action in one background
into the classical master action of the other background, while at the
same time, the antibracket in one background is taken into the
antibracket in the other background. Since a (classical) master action
and an antibracket define a classical theory in the BV formulation, the
field transformation maps one theory into the other. It then follows that
the classical closed string actions (obtained by keeping the ghost number
zero fields of the classical master action) in the two backgrounds are
related by this field redefinition, restricted to the ghost number zero
fields.

If we turn to the complete quantum theory it is less clear what is
the physically significant notion of background independence. Perturbatively,
quantum theories are usually defined only after gauge fixing. Is
background independence therefore a relation between gauge fixed theories ?
or is it a relation between  effective actions ? Moreover, what is the
physical significance of the shift of the field ? We should note that,
while in classical closed string field theory
conformal field theories are consistent backgrounds,
in the quantum theory they need not be.
Quantum closed string field theory
formulated around conformal
theories, contains elementary vertices giving rise to terms linear in the
string field (at higher orders in $\hbar$). Therefore
quantum closed string field theory
is formulated on backgrounds around which the action is
not stationary. We will see in this paper that in the BV approach we
can define a satisfactory notion of background independence of the quantum
theory.

This notion of background independence is based on the idea that
there should be a formal equivalence between the theories formulated
around the two different backgrounds. This formal equivalence,
in turn, is proved by establishing the strict (non-formal) equivalence
of appropriate action weighted measures in the two theories.
This point is relevant even for the classical theory, where
we showed the existence of a diffeomorphism mapping the actions and the
symplectic structures. This is a strict (non-formal) statement. One may
be tempted to conclude that this diffeomorphism proves that the observables
of the two theories agree. This, however, is just a formal statement.
It would be a strict statement for compact systems with a finite number of
degrees of freedom, but is only a formal equivalence for typical field
theories. The reason is that observables involve
functional integrals over noncompact
(infinite dimensional) spaces, and boundary conditions are typically necessary
to make sense out of the integrals. A standard boundary condition is the
asymptotic vanishing of the fields. This condition, however, cannot be
imposed simultaneously on field configurations related by a diffeomorphism
that involves a constant shift. This constant shift, representing the
classical solution moving us from one background to the other is the
obstruction to strict physical equivalence of the two backgrounds. Background
independence is nevertheless a well-defined strict relation between two
theories.

To state this relation explicitly, let us recall that
a quantum theory in the BV formulation, is defined by the data
$(M ,\omega ,d\mu , S)$, where $M$ is the supermanifold of field/antifield
configurations, $\omega$ is an odd symplectic form on $M$, $d\mu$
is a volume element on $M$ leading to a nilpotent $\Delta_{d\mu}$ operator,
and $S$ is the master action, a function on $M$ satisfying the master
equation. For closed string field theory, $M$ is a subspace $\H$ of the
state space of a conformal field theory, the symplectic form $\omega$, the
measure $d\mu$ and the action $S$ are all well-known. The problem of
background independence of quantum closed string field theory
must therefore begin with
two conformal field theories $x$, and $y$ and the data
$(\H_x,\omega_x, d\mu_x, S_x)$ and $(\H_y,\omega_y, d\mu_y, S_y)$.
Since background independence must imply a formal equivalence between
the two string field theories, we must produce a
diffeomorphism between the spaces $\H_x$ and $\H_y$ which establishes
the equivalence. Experience with the classical master action [\senzwiebach]
indicates that the diffeomorphism ought to be symplectic and therefore must
carry $\omega_y$ to $\omega_x$.  A further guide are the results of
Ref.[\hatazwiebach], where it was seen that the symplectic diffeomorphism
relating string field theories using different string
vertices does not preserve the volume element $d\mu$ nor the action $S$.
Such diffeomorphism, however, can be seen to preserve the action weighted
measure $d\mu_S\equiv d\mu e^{2S/\hbar}$.
This result suggests that the symplectic diffeomorphism, relating
string field theories around two different backgrounds, must be required to map
the measure $d\mu_y\, e^{2S_y/\hbar}$ to $d\mu_x\, e^{2S_x/\hbar}$.
This is the condition of quantum background
independence proposed in this paper. Since
the distinction between the path integral measure and the action is
ambiguous, it is not surprising that only a relevant combination of both
is background independent.

The factor of two multiplying the action in $d\mu\, e^{2S/\hbar}$
is not an accident of normalization. This factor is unusual, but correct.
We confirm this in several ways. First of all the object
$d\mu_S = d\mu\, e^{2S/\hbar}$ is  rather fundamental.
It is implicit in [\schwarz] that,
if $d\mu$ leads to a consistent delta operator $\Delta_{d\mu}$ (that is,
a nilpotent one), then the action-weighted measure $d\mu_S$
also leads to a consistent operator $\Delta_{d\mu_S}$ whenever
the action $S$ satisfies the master equation. We also observe that
the definition of observables does not require the separate existence
of $d\mu$ and of $S$, but rather the existence of $d\mu_S$ (suggesting that
there could be theories where only one consistent measure $d\mu_S$ exists).
It has also been puzzling that,
while the classical master equation
implies the existence of gauge transformations for the classical master
action, the quantum master equation only seemed to imply BRST transformations
for the quantum theory. It has now been shown [\senzwiebachgauge]
that the quantum master equation implies
the existence of gauge transformations leaving invariant the measure
$d\mu_S$. These transformations agree with the gauge transformations
of the classical theory in the limit $\hbar\to 0$.
Finally, the main reason why $d\mu_S$ is the correct background independent
object is that it guarantees the formal background independence of observables.
In BV theory, observables arise by integration of suitable measures over a
Lagrangian submanifold $L$ of the full manifold $M$ of field/antifield
configurations.  The dimension of the lagrangian submanifold $L$ is half of
that of $M$, and the measure $d\mu$ on $M$ induces a measure $d\lambda$ on
$L$ by an operation involving a square root [\schwarz]. It then follows that
the integrals over $L$ defining observables use the measure
$d\lambda_S \equiv d\lambda \,e^{S/\hbar}$. The symplectic diffeomorphism,
if it maps $(d\mu_S)_x$ and $(d\mu_S)_y$ into each other, will map
$(d\lambda_S)_x$ and $(d\lambda_S)_y$ into each other, and the
(formal) background independence of observables will follow.
The main objective of the present paper is the construction
of the symplectic diffeomorphism implementing the background independence
of quantum CSFT. We will present this construction
for the case of backgrounds
corresponding to nearby conformal field theories related by an exactly
marginal perturbation.

Our success in establishing and proving
a criterion for quantum background independence based on the
master action provides further evidence of
the deep significance of the BV formulation of string theory.
A clear proof of background independence is of value, not only
because the absence of background independence would be a catastrophe,
but because such a proof explains how background independence is realized.
Such understanding is likely to be necessary for further progress.
For example, the background independence of $d\mu_S$
suggests that this is the object we should
try to construct (on theory space) to achieve a manifestly
background independent formulation of string field theory.

Our proof of background independence required completing the construction
of the closed string field theory master action. The condition that
the master action satisfies the master equation fixes this action up
to field independent constants. This implies that when we construct a single
string field theory, the measure $d\mu_S$ is only fixed up to a constant.
This time, however, we are constructing string field
theories over a CFT theory space, and therefore we must compare
measures $(d\mu_S)_x$ at different points $x$. The normalization
of these measures has to be fixed consistently if one is to have
background independence. In the spirit of our analysis, where the
string action is expanded as $S\sim\sum_{g,n} S_{g,n} \hbar^g$,
the overall constant also has an $\hbar$ expansion. We show that
the $\hbar$ dependent terms arise from the vacuum string vertices $\V_{g,0}$
(with $g\geq 2$) which in turn, are defined by the geometrical recursion
relations of string field theory. Such role for vacuum vertices was
anticipated in Ref.[\zwiebachlong] where it was also pointed out that
the minimal area methods are applicable to vacuum graphs. The case
of the $\hbar$ independent overall constant is quite fascinating. When
we choose basis sections $\ket{\Phi_i}_x$ for $\H_x$
throughout theory space, the
measures $d\mu_x$ can be written as $d\mu_x = \rho(x)\prod_i d\psi^i_x$,
where $\psi^i$ denotes the coordinate associated to the basis vector
$\ket{\Phi_i}$. In addition to $\rho(x)$, the term $S_{1,0}$ also affects
the $\hbar$-independent part of the normalization of the measure $d\mu_S$.
We show that the conditions of
background independence can be satisfied if $\rho(x)$ is chosen to
satisfy the equation ${\p \ln\rho\over\p x^\mu} = \hbox{str}\,\wt\Gamma_\mu$,
where $\wt\Gamma_\mu$ is a connection on theory space, and $S_{1,0}$ is
the integral of the one loop CFT partition function over part of the
moduli space of tori. We find it thought provoking
that some information about
the connection is necessary to formulate a background independent
action for string field theory. It suggests
that a connection on theory space is more than just a technical tool
to prove background independence of string field theory. We feel that
the appearance of a connection in the formulation of quantum string field
theory suggests that a dynamical
connection might be necessary to achieve a manifestly
background independent formulation of string field theory.

The proof of local background independence presented in [\senzwiebach]
for classical closed string field theory made special use of the
particular properties of polyhedra.
Since the classical part of the quantum master
action for closed string field theory cannot be built with polyhedra (
it requires stubs), the ability to deal with general string vertices
was necessary to address quantum background independence.
The present formalism will achieve this goal. In doing so we reach
the conclusion that background independence is simply
a consequence of the geometrical consistency conditions
of the string vertices (however they might be chosen)
and of the existence of a theory space with a connection.

The main development that paved the way to a geometrical proof of
background independence
was the setting up of a
formalism to deal efficiently with subspaces of moduli
spaces of Riemann surfaces. In fact we introduce a complete BV  structure
acting on subspaces of moduli spaces.
We introduce an antibracket  $\{ \, ,\, \}$ that, acting on
two spaces of Riemann surfaces, produces a third space whose elements
are all the
surfaces obtained by twist-sewing two surfaces each in one of the
original spaces. We also introduce an operator $\Delta$, which, acting
on a space of surfaces, gives a space of surfaces
whose elements are all the surfaces
obtained by twist-sewing two punctures
in a surface in the original space.
Defining carefully the orientation of the subspaces involved, this
antibracket and delta operator are shown to satisfy all the formal
properties associated to these objects in BV quantization. This
correspondence is not accidental, for we show that these operations
on moduli spaces are indeed represented by the BV antibracket and
delta operator at the level of functions on $\H$.
Finally, we introduce an operator $\K$ which, applied to a space of surfaces,
gives us a space of surfaces whose elements are surfaces in the original
set but with one additional puncture.
This operator is intimately related to
the computation of covariant derivatives of surface states using
the special connection $\wh\Gamma_\mu$ [\kugozwiebach,\rangaconnection,
\rangasonodazw].

In analogy to the case of the classical theory, the field redefinition
relating the two quantum string field theories is defined by
vertices $\B_{g,n}$ that interpolate between the string vertices
$\V_{g,n}$ and a new set of vertices $\V'_{g,n}$. A $\V'$ vertex
contains all the surfaces obtained in three possible ways;
surfaces arising from $\K$ acting on a $\V$ vertex,
from $\Delta$ acting on a $\B$ vertex,
or by twist sewing of a lower dimensional
$\B$ vertex to a lower dimensional $\V$ vertex.

In a recent paper, Witten [\wittenqbi] has discussed a problem of quantum
background independence in the context of superconformal field theories
representing Calabi-Yau backgrounds. In this context, in addition
to parameters representing true changes of backgrounds, there is
an extra set of parameters representing BRST-trivial deformations, that,
through an anomaly, end up creating true deformations [\ooguri ].
Further discussion of the obstructions to full background independence
in this context has
been given in Ref.[\oogurii], where the relevant closed string field
theory was constructed.
The methods of our present work may be useful to investigate if strict
background independence can be achieved.

Since we have introduced the analog of anti-bracket and the $\Delta$
operator in the moduli space of Riemann surfaces, it is a natural question
to ask if we can also find an operator $(\cdot)$ in the moduli space, that
will be an analog of ordinary multiplication in the space of functions. It
turns out that such an operation can indeed be defined, but we need to
generalize the space by including moduli spaces of surfaces that can
contain disconnected components. Furthermore, in this new space the string
field theory vertices can be shown to generate a non-trivial cohomology
element of a nilpotent operator $(\p +\hbar\Delta)$, where $\p$ denotes
the operation of taking the boundary of a space.

Let us sketch briefly the contents of the present paper. In \S2
we develop the techniques to deal efficiently with
spaces of surfaces by introducing the operations $\{\,,\,\}$, $\Delta$,
and $\K$ which act on subspaces of moduli spaces. This section is
purely geometrical. In \S3
we review the construction of the volume element and delta operator suitable
for closed string field theory, as well as the construction of the quantum
master action. We show how the operators introduced for moduli spaces,
are represented in $\H$ by the standard BV antibracket and delta operators.
In \S4 we define our conditions for quantum background independence,
discuss their physical and geometrical significance, and obtain the
explicit form of the conditions for nearby backgrounds. Our proof of
quantum background independence of the field dependent, as well as field
independent but $\hbar$ dependent part of the action
weighted measure is given in \S5. We prove in \S6 the
background independence of the $\hbar$ independent normalization of
the measure $d\mu_S$. In \S7 we introduce the $(\cdot)$ operation on
the moduli space of Riemann surfaces with disconected components, and
rewrite the recursion relations of closed string field vertices as a
cohomology condition in this  space. We also reinterpret the
recursion relations that determine the $\B$ spaces.
We offer some concluding observations and remarks in \S8.

\chapter{Operations on Spaces of Riemann Surfaces}

In the present section we introduce notation that
will enable us to manipulate with ease spaces of Riemann surfaces.
The spaces of surfaces we have in mind are made of formal sums
(in the sense of homology) of oriented subspaces of moduli spaces.
Given two spaces
of surfaces $\A_1$ and $\A_2$,
we define a third space of surfaces $\{ \A_1 , \A_2\}$, whose
surfaces are obtained by sewing surfaces of the original two spaces.
This sewing is done with sewing parameter $t=e^{i\theta}$ with
$\theta \in  [0,2\pi]$ (twist-sewing). We discuss the properties of
this bracket operation on spaces of surfaces, particularly with respect
to the operation $\p$ of taking the boundary of a space of surfaces.
We also define an operation $\Delta$, which acting on a space of surfaces,
gives a new space of surfaces
whose elements are obtained by twist-sewing two punctures in each surface
of the original space. Finally, we introduce the operation $\K$, which
acting on a space of surfaces, gives us a space of surfaces whose elements
have one additional puncture. Our use of a bracket $\{ \, , \, \}$, and
$\Delta$ to denote operations on spaces of surfaces is not accidental.
These operations satisfy the formal properties expected from these
objects in standard BV quantization. As we will see in \S3,
the action of the BV antibracket and delta operator on functions
in the vector space $\H$ are, in fact, represented by the corresponding
operations in moduli space.
In \S7 we complete the construction
of a BV algebra on spaces of surfaces by introducing the relevant
dot product.

\section{Antibracket for Moduli Spaces}

Let $\wh\P_{g,n}$ denote, as usual, the moduli space of Riemann surfaces
of genus $g$ with $n$ punctures, with a choice of a local coordinate, up to a
phase, around each puncture. This space has the structure of a fiber bundle
with base space the moduli
space $\M_{g,n}$ of genus $g$ Riemann surfaces with $n$ punctures (without
a choice of local coordinates). For any surface $\Sigma\in \M_{g,n}$, the
fiber over $\Sigma$ is the set of all points of $\wh\P_{g,n}$ representing the
surface $\Sigma$ equipped with some choice of local coordinates at its
punctures. In the present section we will develop the notation and basic
results that will allow us to manipulate with ease subspaces of
$\wh\P_{g,n}$. In
particular we will study what happens when we construct new subspaces by
sewing operations.

Given two subspaces $\A_1\subset\wh\P_{g_1, n_1}$ and
$\A_2\subset\wh \P_{g_2,n_2}$, we select a {\it fixed} labelled puncture $P_1$
on all surfaces of $\A_1$ (one out of $n_1$ possible choices),
and a {\it fixed}
labelled puncture $P_2$ on all surfaces of $\A_2$
(one out of $n_2$ possible choices).
We then define the subspace $\{ \A_1 ,  \A_2\}' \subset\wh
\P_{g_1+g_2,n_1+n_2-2}$
as the set of surfaces obtained by twist sewing every surface in $\A_1$
to every surface in $\A_2$ using the selected punctures
$P_1$ and $P_2$.
If $\A_1$ is a subspace of dimensionality $d_1$,
and $\A_2$ is a subspace of dimensionality $d_2$, then $\{\,\A_1 , \,\A_2 \}'$
is a subspace of dimensionality $d_1+d_2 +1$, where the one extra
dimension arises from the twist angle.
The bracket $\{\, , \, \}'$ depends, in general, on the choice
of labelled punctures, and that is why we have introduced the prime.
If a subspace $\A$ is symmetric (in the sense of assignment of labelled
punctures [\senzwiebach]) then the choice
of a labelled puncture for sewing is irrelevant. Therefore the bracket
of two symmetric subspaces is uniquely defined without specifying punctures.
For this case we will introduce later an unprimed bracket, where in addition,
the resulting space of surfaces will be symmetric.
While only the
unprimed bracket will be necessary for our applications,
the primed bracket is a more basic object.

We need to define the orientation of the subspace $\{\A_1, \A_2\}'$.
To this end we must define an ordered set of basis vectors
$[\cdots ]$ for the
tangent space to $\{ \A_1 , \A_2\}'$ at any point. Consider a surface
$\Sigma \in \{\A_1 ,  \A_2\}' $ obtained by sewing a surface $\Sigma_1\in \A_1$
to a surface $\Sigma_2\in \A_2$. Let $[ \A_1 ]$ denote the orientation
of $T_{\Sigma_1}\A_1$ and  $[ \A_2 ]$ denote the orientation
of $T_{\Sigma_2}\A_2$. Each basis vector in $[\A_1]$ or $[\A_2]$ defines
naturally a basis vector in $T_{\Sigma} \{\A_1, \A_2\}'$, since any deformation
of the surfaces to be sewn determines a deformation of the sewn surface
(when we keep the sewing parameter fixed).\foot{Strictly speaking,
since the surfaces in $\A_1$ and $\A_2$ are equipped with local
coordinates {\it up to a phase}, deforming these surfaces keeping the
sewing parameter fixed is not a well defined concept. Thus the set of
vectors $\{\A_1\}$ and $\{\A_2\}$ are defined up to addition of
vectors proportional to ${\p\over \p\theta}$. This ambiguity does not
affect the orientation defined by $[\{\A_1\}, {\p\over\p\theta},
\{\A_2\}]$.}
Let $\{ \A_1\}$ denote the
set of basis vectors in $T_\Sigma \{\A_1,\A_2\}'$ arising from $[\A_1]$,
and $\{ \A_2 \}$  denote the set of basis vectors in
$T_\Sigma \{\A_1, \A_2\}'$ arising from $[\A_2]$. We define the
orientation of $T_\Sigma \{\A_1,\A_2\}'$ as
$[\, \{\A_1\}, {\p\over\p\theta},\{\A_2\}\, ]$, where
${\p\over\p\theta}$ is the tangent vector associated to changes in
the sewing twist angle $\theta$, arising in $z_1 z_2 = \exp (i\theta)$,
whenever we sew together the punctures associated to the local coordinates
$z_1$ and $z_2$.\foot{In ref.[\senzwiebach] the bracket of two spaces
of surfaces was indicated by the operation $\times$. In defining the
orientation of  $\B\times\V$, we had taken the ordering to be
$[\, \{\B\}, \{\V\},{\p\over \p\theta}]$. Since the manifold
$\V$  was always even dimensional, the present bracket agrees
with $\times$ in the cases relevant to  ref.[\senzwiebach].}

For any region $\A\subset \wh\P_{g,n}\,$, $\p\A$ will denote the boundary
of $\A$. The orientation of $\A$ induces an orientation on $\p\A$ as
usual. Given a point $p\in \p\A$, a set of basis vectors $[v_1, \cdots v_k]$
of $T_p(\p\A)$ defines the orientation of $\p\A$ if $[\,n,v_1, \cdots v_k\,]$,
with $n$ a basis vector of $T_p\A$ pointing outwards,\foot{To obtain
an outward vector one constructs a diffeomorphism between the neighborhood
of $p$ and a suitable half-space. The outward vector is the image under
the diffeomorphism of the standard normal to the half space (see, for
example [\spivak]).}
is the orientation of $\A$ at $p$. From this definition it is clear that
$$ \p\, \{\A_1 ,  \A_2\}'  = \{ \p\A_1 ,  \A_2\}' + (-)^{\A_1 +1} \{ \A_1 ,
\p\A_2 \}' \, .\eqn\ebountimesx$$
The factor of $(-)^{\A_1+1}$\foot{$(-)^\A$ stands for
$(-)^{\hbox{dim}\A}$, where dim$\A$ is the real dimension of $\A$.} in
the second term arises because in order to compare the orientation of
$\p \{ \A_1, \A_2 \}'$, when $\A_2$ is at a boundary, with that of
$\{ \A_1 ,  \p\A_2\}'$ one must move the outward vector across the
tangent vector $\p/\p\theta$, and across all the tangent vectors of $\A_1$.
Therefore, $\p$ acts as an odd derivation of the bracket,
with a space of surfaces treated as an even object if it is even dimensional,
and as an odd object if it is odd dimensional.
It then follows from these rules, and the definition of the bracket that
$$\{\A_1 , \A_2\}'= -\,(-)^{(\A_1+1)(\A_2+1)}\,\{\A_2,  \A_1 \}'\,.\eqn\comxx$$
This indeed coincides with the conventional exchange property of the
BV antibracket.
It is convenient to introduce a notation for sewing of more than two
spaces of surfaces. We will denote by $\{ \A_1, \A_2, \A_3\}'$
the space of surfaces whose elements are obtained
as a result of twist-sewing a puncture of each surface $\Sigma_1$ in $\A_1$
to a puncture of each surface $\Sigma_2\in \A_2$, and another puncture of
$\Sigma_2$ to a puncture of each surface $\Sigma_3$ in $\A_3$.
We then have that
$$\{\A_1, \A_2, \A_3\}' = -\,(-)^{\A_1\A_2 +\A_2\A_3 +\A_3\A_1 }\,
\{\A_3 , \A_2 , \A_1\}' \, ,\eqn\ecrossexchangex$$
where the extra minus
sign on the right hand side appears due to the
exchange of the two $\p/\p\theta$'s. Since they are associated to
tangent vectors corresponding to different sewing parameters, they
must be moved through each other. Another relation that follows
naturally from the above
discussion and will be useful to us is,
$$ \bigl\{ \{\A_1, \A_2\}' , \A_3 \bigr\}'= \cases{
\{\A_1, \A_2 , \A_3\}'\,, \cr
-(-)^{(\A_1+1)(\A_2+1)} \{ \A_2 ,\A_1, \A_3\}'\,, }
\eqn\emultiproducta$$
according to whether $\Sigma_3\in\A_3$ is sewn to a puncture on
$\Sigma_2\in\A_2$ (first case) or to a puncture on
$\Sigma_1\in\A_1$ (second case).

As mentioned before, if the spaces $\A_i$ are symmetric the primed bracket
does not need the specification of the punctures to be sewn. The resulting
space of surfaces, however, will not be symmetric in general. Let {\bf S}
denote a symmetrization operator, and introduce the unprimed bracket as
$$ \{ \A_1 , \A_2 \} \equiv\, {\bf S}
\,\{ \A_1 , \A_2 \}'\,\, .\eqn\rbracket$$
When the resulting space of surfaces has $n$ labelled punctures, the
operator ${\bf S}$ adds the result of the primed bracket, for all possible
ways of splitting the $n$ labelled punctures among the two initial spaces
of surfaces. Similarly, we define
$$ \{ \A_1 , \A_2 ,\A_3 \} \equiv\, {\bf S} \,\{ \A_1 , \A_2, \A_3 \}'\,\, ,
\eqn\rmbracket$$
where, if the resulting space of surfaces has $n$ labelled punctures,
the operator ${\bf S}$ adds the result of the primed bracket, for all possible
ways of splitting the $n$ labelled punctures among the three initial spaces
of surfaces. If we use the unprimed bracket Eqn.\emultiproducta\ need not deal
with separate cases. It becomes
$$ \Bigl\{ \{ \A_1 , \A_2\}\,  , \, \A_3\Bigr\} =
\, \{\A_1, \A_2 , \A_3\}
-(-)^{(\A_1+1)(\A_2+1)}\, \{\A_2 , \A_1, \A_3\}\, .\eqn\emultiproduct$$
The previous identities satisfied by the primed bracket immediately
imply the identities
$$ \p\, \{\A_1 ,  \A_2\}  = \{ \p\A_1 ,  \A_2\} + (-)^{\A_1 +1} \{ \A_1 ,
\p\A_2 \} \, ,\eqn\ebountimes$$
$$\{\A_1 , \A_2\}= -\,(-)^{(\A_1+1)(\A_2+1)} \,\{\A_2,  \A_1 \}\,,\eqn\comx$$
$$\{\A_1, \A_2, \A_3\} = -\,(-)^{\A_1\A_2 +\A_2\A_3 +\A_3\A_1 }\,
\{\A_3 , \A_2 , \A_1\} \, .\eqn\ecrossexchange$$
As befits an antibracket, the Jacobi identity indeed holds,
$$ (-)^{(A_1+1)(A_3+1)} \Bigl\{ \{ \A_1 , \A_2\}\,  , \, \A_3\Bigr\}\,+\,
\hbox{cyclic} \, = 0\,,\eqn\jcbdntty$$
as can be verified using \emultiproduct\ and \ecrossexchange. We have
therefore defined a consistent antibracket acting on symmetric subspaces
of moduli spaces of Riemann surfaces. There is also a delta operator
on moduli space satisfying all requisite properties. This operator is
the subject of next subsection.

\section{Delta Operator on Moduli Spaces}

In our analysis we shall also need an operator $\Delta$.  Acting on
a subspace $\A\subset\wh\P_{g,n}$, the operator $\Delta$
will be defined to give a subspace
$\Delta\A \subset\wh\P_{g+1,n-2}$,
representing the surfaces obtained by twist sewing two fixed-label punctures
for every surface in $\A$. The result $\Delta\A$ depends on the choice of
labels, unless the subspace $\A$ is symmetric.
Since in our analysis the vertex $\A$ will always be symmetric in the
punctures that are sewed by the $\Delta$ operator, we shall include an explicit
factor of $1/2$ in the definition of $\Delta \A$ in order to avoid double
counting a given contribution to $\Delta \A$ arising from the interchange
of the two punctures.  \foot{Note that this
convention may leave an explicit factor of 1/2 in the definition of
$\Delta\A_{0,3}\subset \wh\P_{1,1}$. This is the case, for example,
when $\A_{0,3}\in\wh\P_{0,3}$ contains a single point in $\wh\P_{0,3}$.
Thus $\Delta\A_{0,3}$ is typically a formal space of surfaces.
When doing integrals this multiplicative
factor in the definition of the space is simply converted to a multiplicative
factor for the integrand. }
The orientation of $\Delta\A$ is defined to be
$[ \p/\p\theta,\{\A\} \,]$. The operator $\Delta$ is an odd
operator, just as the corresponding object in BV quantization,
and satisfies the following properties:
$$\Delta\p\A=-\p\Delta\A, \eqn\edeltaboundary $$
$$\Delta \, \{ \A_1,  \A_2\}' = \cases{
\{ \Delta\A_1, \A_2 \}'\, ,
\cr (-)^{(\A_1+1)}\, \{ \A_1, \Delta\A_2\}'\, ,
\cr (-)^{\A_1} \A_1\asymp \A_2 \,, \cr} \eqn\edeltatimes $$
depending on whether $\Delta$ sews two punctures of $\A_1$ (first case),
two punctures of $\A_2$ (second case) or one puncture of $\A_1$ with
one puncture of $\A_2$ (third case).
In the last term, two of the punctures of $\A_1$ are sewed to two of
the punctures of $\A_2$.
For this term, the ordering of the tangent vectors is
$[\{\A_1\}, {\p\over\p\theta_\Delta}, {\p\over\p\theta_{ab}}, \{\A_2\}]$,
where $\theta_\Delta$ is the sewing parameter associated with the $\Delta$
operator and $\theta_{ab}$ is the sewing parameter associated with the
antibracket on the left hand side.

In all of our analysis, the vertices $\A_i$ that we shall encounter will
be symmetric in the relevant legs,
namely, the ones which are sewn by the bracket or the $\Delta$ operator.
If $\A_1$ and $\A_2$ in Eqn.\edeltatimes\
denote symmetric vertices, then
the last term on the right hand side of this equation vanishes.
To see this let us denote the pairs
of {\it unlabelled} punctures that are sewn
by $(P_1, P_2)$ and $(Q_1, Q_2)$, with $P_1, Q_1$ lying on a surface
$\Sigma_1\in \A_1$
and $P_2, Q_2$ lying on a surface $\Sigma_2\in\A_2$.
Moreover, let $\p/\p\theta_P$ be the tangent vector associated with the
sewing of the $P$ punctures,
and $\p/\p\theta_Q$ be the tangent vector associated with the sewing
of the $Q$ punctures.
Assume now that the labelling of the punctures is such that
the surfaces  $\Sigma = \Delta \{\Sigma_1, \Sigma_2\}'$
are constructed with the antibracket sewing
$P_1$ to $P_2$, and $\Delta$ sewing $Q_1$ to $Q_2$.( Recall that
bracket and delta must sew punctures of fixed labels.)
The orientation of the space at $\Sigma$, as explained above,
will contain the tangent
vectors $[ {\p\over \p\theta_Q} ,  {\p\over \p\theta_P }]$ in this order.
The symmetry of the spaces $\A_1$ and $\A_2$ implies
that $\Delta \{ \A_1, \A_2\}'$ contains
surfaces $\Sigma' =\Delta \{ \Sigma'_1, \Sigma'_2 \}'$ where
$\Sigma'_1$ differs from $\Sigma_1$ only by the exchange of the labels
on the fixed punctures $P_1$ and $Q_1$,
and $\Sigma'_2$ differs from $\Sigma_2$
only by the exchange of the labels on the fixed punctures $P_2$ and $Q_2$.
This time the antibracket will end up sewing the $Q$ punctures, and
$\Delta$ will sew the $P$ punctures. The resulting
$\Sigma'$ will contain the same surfaces as $\Sigma$ but this time
the orientation will contain the sewing tangent vectors in the opposite order
$[ {\p\over\p\theta_P} , {\p\over \p\theta_Q}]$.
Therefore $\Sigma$ and $\Sigma'$ contribute
to $\Delta \{ \A_1, \A_2\}'$ with opposite orientations and hence cancel.
This gives, for symmetric vertices, and with the unprimed bracket
$$ \Delta  \{ \A_1, \A_2\}=\,\,\{\Delta\A_1,
\A_2\}+(-)^{\A_1+1} \{\A_1,\Delta\A_2\}\,\,.\eqn\edeltatisym$$
This is in correspondence to the fact that in BV quantization, the delta
operator defines a derivation of the antibracket.
The same symmetry argument shows that for a symmetric vertex $\A$,
$$\Delta(\Delta\A)= \Delta^2 \A = 0 \,,\eqn\edeltasquare$$
due to the antisymmetry under the exchange of the tangent vectors associated
with the two sewing angles. This result is again in correspondence with
the nilpotency of the delta operator in BV quantization.

For our analysis, it will be useful to introduce the complex
$$\wh\P = \oplus_{g,n} \wh\P_{g,n} \quad \hbox{with} \,
\cases{n\geq 3 \,\,\hbox{for}\,\,g=0,\cr  n\geq 1 \,\,\hbox{for}\,\,
g=1\cr n\geq 0 \,\,\hbox{for}\,\, g\geq 2\,.}\eqn\edefhatp $$
whose elements are formal sums of the form $\sum_{g,n} a_{g,n}\A_{g,n}$,
where $a_{g,n}$ are real numbers
and $\A_{g,n}\subset \wh\P_{g,n}$ are subspaces which do not include
surfaces arbitrarily close to degeneration. Here $\wh\P_{g,0} \equiv \M_{g,0}$,
as we have no punctures.
The definition of the $\Delta$ and the $\{\, , \, \}$ operators is
extended to this complex by treating them as linear and bilinear
operators respectively:
$$\Delta (a_1 \A_1 + a_2 \A_2) = a_1 \Delta\A_1 + a_2
\Delta\A_2 \,, \eqn\edeltaextension $$
$$\eqalign{ \{\, a_1 \A_1 + a_2 \A_2\,, \,b_1 \B_1 + b_2 \B_2\, \}  &=
\, \,a_1 b_1 \{\A_1, \B_1\}
+ a_1 b_2 \{\A_1, \B_2\}\cr & \,\,\,+ a_2 b_1 \{\A_2, \B_1\} +
a_2 b_2 \{\A_2, \B_2\}\,.
\cr}\eqn\ebracketextension$$

The above notation will now be used to write down the geometrical
recursion relations of closed string field theory.
The closed string field theory vertices $\V_{g,n}$ ($n\ge 3$ for $g=0$,
$n\ge 1$ for $g=1$, $n\ge 0$ for $g\ge 2$) are $d_{g,n}=(6g+2n-6)$
dimensional subspaces of $\wh\P_{g,n}$ and satisfy the recursion
relations [\sonodazwiebach,\zwiebachlong]
$$\eqalign{
\p\V_{g,n} &=- \half\sum_{g_1+g_2=g\atop n_1+n_2 = n+2}
\hskip-6pt\bigl\{ \V_{g_1,n_1}\, , \, \V_{g_2, n_2}\bigr\}\, -\,
\Delta \V_{g-1, n+2}\, ,\cr
&=- \half\,\sum_{g_1=0}^g \sum_{m=1}^{n+1}
\bigl\{ \V_{g-g_1, n-m+2}\,,\, \V_{g_1, m}\bigr\} - \Delta \V_{g-1, n+2}\, ,
}\eqn\propvgn$$
In writing the above equation we have defined
$$\V_{0,n} \equiv 0 \quad \hbox{for}\,\,n\leq 2\, , \eqn\fconv$$
otherwise the range of summation over $m$ will depend on $g$. This definition
is consistent with \propvgn\ since, for $n\leq 2$, $\p\V_{0,n}$ computed
with the help of this equation is indeed zero. Although the form
of the field independent part of the action, given by $\V_{g,0}$, is not
specified by the master equation, we shall take $\V_{g,0}$ for $g\ge 2$ to
satisfy the above equation[\zwiebachlong]. The case of $g=1, n=0$ is
special, and will be discussed in \S6.

We now define a special vector in the complex $\wh\P$. We define
$$\V \equiv  \sum_{ g,n}\hbar^g \V_{g,n}  \quad\hbox{with}\quad
\cases{n\geq 3 \,\,\hbox{for}\,\,g=0 ,\cr
n\geq 1\,\,\hbox{for}\,\,
g=1,\cr n\geq 0 \,\, \hbox{for}\,\, g\ge 2\,.}\eqn\dfnvsp$$
Note that we have included the vacuum vertices $\V_{g,0}$ for $g\ge 2$,
but not a genus one vacuum vertex.
It then follows from \propvgn\ that the recursion relations can be written as
$$\p \V   + \hbar \Delta \V + \half \{ \V , \V \} = 0 \, .\eqn\recrelnew$$
An even nicer reformulation of the recursion relations will be given
in \S7.

\section{The Operator $\K$ on Moduli Spaces}

Consider a genus $g$ Riemann surface $\Sigma$ with $n$ punctures
and equipped with local coordinates determined by coordinate curves
around the punctures. We consider the general case when $g$ and $n$
can take arbitrary values. Assume also that the coordinate curves
do not intersect, and therefore $\Sigma$ minus the unit
disks $D_i^{(1)}$ (bounded
by the coordinate curves) is a nonvanishing region of the
surface.
We now define a subspace $\K (\Sigma) \in \wh\P_{g,n+1}$ which will
contain the $n+1$ punctured surfaces corresponding to $\Sigma$ with
an extra puncture lying anywhere on the region
$\Sigma-\cup_i D_i^{(1)}$.
If the surface $\Sigma$ has no punctures, the extra puncture can lie
anywhere on it. The coordinate at the extra puncture
will be fixed arbitrarily, but continuously as it moves on the surface.
It follows that $\K(\Sigma)$ is a section over the subspace of
$\M_{g,n+1}$ defined by the surface $\Sigma$ with the extra
puncture lying somewhere on the surface minus its unit disks.
This section is two dimensional, and its orientation is defined
by $[V(v_1),V(v_2)]$, where $[v_1,v_2]$ define the standard orientation
of $\Sigma$, and $V(v)$ denotes the tangent (in $\wh\P_{g,n+1}$)
representing
the deformation induced by moving the extra puncture in the direction
indicated by $v$. While we have said that the coordinate at the extra
puncture can be fixed arbitrarily, it will be useful to fix it explicitly.
One way to do it is as follows.
On $\Sigma$ we find the minimal area metric,
satisfying the requirement that all homotopically
non-trivial closed curves on the surface have length
$\ge 2\pi$ [\zwiebachlong]. The coordinate curves
are then some curves $C_i$ on the surface (that need not correspond
to critical trajectories nor geodesics).
Then at every point $p$ on
$\Sigma$ minus the disks bounded by $C_i$ we shall take the coordinate curve
associated with the puncture at $p$ as the locus of points at a distance $a$
from $p$. If $a<\pi$ this rule determines well defined coordinate disks
that vary continuously as $p$ moves on $\Sigma$.

Given a subspace $\A\subset\wh\P_{g,n}$ we define $\K\A$
to be the subspace of $\wh\P_{g,n+1}$ representing the surfaces
$\K(\Sigma)$ for all surfaces $\Sigma \in \A$. The space $\K\A$
is connected if $\A$ is, if the local coordinate
at the extra puncture changes continuously as we move in $\A$.
This condition is automatically satisfied if the local coordinate at the
extra puncture is chosen using the prescription given at the end of the
last paragraph.
Let $[\A]$ denote the
orientation of $\A$, and $\{ \A \}$ denote the tangent vectors
on $\wh\P_{g,n+1}$ arising from the tangent vectors in $[\A ]$.
(These are defined up to the addition of linear combinations of the vectors
$V(v_1)$, $V(v_2)$.)
The orientation of $\K \A$
is defined to be $[V(v_1),V(v_2),\{\A\}\,]$, where, again, $[v_1,v_2]$
is the orientation of the Riemann surface.

We shall now introduce a new
symbol $\approx$ to express relations that hold up to
the local coordinate at the extra puncture.
More precisely, we say $\A_{g,n+1} \approx \B_{g,n+1}$ if
$\pi_f (\A_{g,n+1}) = \pi_f (\B_{g,n+1})$,
where $\pi_f$ is the projection map that forgets about the coordinate
at the special puncture, that is, $\pi_f$ is
the map from $\wh\P_{g,n+1}$ to the
space $\wh\P_{g,n,1}$ of surfaces of genus $g$ and $n+1$ punctures,
with local coordinates up to phases around $n$ of the punctures.
For a space $\A$ with a special puncture, and having
a tangent vector representing a deformation of the coordinate
at the special puncture, we define $\pi_f (\A)=0$.
The space $\pi_f (\A)$, if nonvanishing will therefore
have the same dimensionality of $\A$. In this case, the
orientation will be defined by
$[\pi_f (\wh V_1),\cdots\,,\pi_f (\wh V_n)]$,
where $[\wh V_1,\cdots, \wh V_n]$ define the orientation of $\A$.

The following properties are easily verified
$$\K\, (\,\{ \A_1 , \A_2\}\, )\, \approx\, \{ \K\A_1 , \A_2\} \,+\,
\{ \A_1 , \, \K \A_2 \}\, , \eqn\distrk$$
$$\K \, (\Delta \, \A) \approx \Delta\, (\K \A)\, .\eqn\comdelta$$
The above equations state that $\K$ is an even derivation of the antibracket,
and it commutes with the $\Delta$ operator.
In these equations the special
punctures created by $\K$ are {\it never} used for sewing; one must think of
these punctures as ones where we insert a marginal operator.
The above equations, with equality up to local coordinates at
the special puncture, is all we shall need for carrying out our
proof of background independence. Nevertheless,
our definition of the local coordinate via minimal area metrics, plus the
compatibility of minimal area with sewing, guarantees that for subspaces
$\A$ of minimal area sections [\zwiebachlong] in $\wh\P$, the above equations
hold strictly (including the local coordinate at the extra puncture).
An important result gives us the boundary of the space
$\K \A$
$$\p\, \bigl( \K \A \bigr) \approx \K \bigl( \p \A \bigr) -
\,\{ \V'_{0,3}\, , \A \}\, ,\eqn\pcomk$$
or more abstractly, as the operator equation
$$[\,\partial\,,\,\K\,]\,\approx\,-\,\{\, \V'_{0,3}\, , \,\, \}\,,\eqn\xz$$
where the left hand side denotes the standard commutator.
We must explain our notation in \pcomk. The second term in the right hand side
involves twist sewing the three punctured sphere $\V'_{0,3}$
to the surfaces in $\A$. This three punctured sphere
$\V'_{0,3}\equiv\V'_3$, introduced in [\senzwiebach],
is symmetric under the exchange
of two punctures, but does not have any further exchange symmetry.
The third, and special puncture of $\V'_3$, is the puncture associated
with the operation $\K$ in the above equation.
It is a puncture that
stands on a different footing from all other punctures, since a
marginal operator is always to be inserted there.
Whenever our antibracket, whose definition included symmetrization,
contains a surface with a special puncture,
that puncture {\it will not be symmetrized over}. That puncture cannot
be used for sewing, and must be considered as if it was filled.
The bracket $\{ \V'_{0,3}\, , \A \}$ effectively sews $\V'_{0,3}$
to each of the punctures of each of the surfaces in $\A$.

This identity arises as follows.
The boundary of $\K\A$ arises from two sources. One is the boundary of
$\A$ itself. This contribution is given by operating $\K$ on $\p\A$
(first term on the right hand side  of Eqn.\pcomk). The second contribution
comes from the boundary of the region of integration over the location of
the new puncture $p$. This is given by configurations where $p$ is
inserted on the coordinate curves $\C_i$ of the original set of punctures.
If we disregard the choice of the local coordinate system at the
special puncture on $\K\A$,
these contributions are given by twist sewing the vertex $\V'_{0,3}$ to
the original surfaces in  $\A$ at all punctures. This gives rise to the
second set of terms on the right hand side of Eqn.\pcomk.
The minus sign arises as follows. When twist sewing $\V'_{0,3}$ to a surface
$\Sigma$ in $\A$ the extra puncture ends up traveling counterclockwise along
each of the coordinate curves of $\Sigma$. The boundary
of the region of integration $\Sigma -\cup_i D^{(1)}_i$, however, gives the
coordinate curves of $\Sigma$, with an orientation corresponding to
clockwise travel.

For subspaces of minimal area sections, the above equation
holds strictly if we define the local coordinate at the special puncture
of $\V'_{0,3}$ in a suitable way. This is done as follows. $\V'_{0,3}$
is mapped to an infinite cylinder of circumference $2\pi$ with the symmetric
punctures at the two points at infinity. The local coordinates at those
punctures are fixed by taking as a common coordinate curve an arbitrary
geodesic circle on the cylinder. On that circle we fix a point to be the
special puncture, and define its coordinate disk to be the set of points
at a distance $\leq a$. This can be seen to yield the same local coordinate
on the special puncture on the surfaces in $\{ \V'_{0,3} ,  \A\}$
as the construction of $\K$ does when the special
puncture lies at the boundary of the region of integration.

For spaces of overlap surfaces
(spaces where each surface $\Sigma$ is equal to the disjoint union of the
unit disks around the punctures)
the identity  in \pcomk\ reduces to
$\{ \V'_{0,3}\, , \A \} \approx 0$, by virtue of $\K \A =0$.
This identity, obtained in [\senzwiebach]
(where instead of using a bracket, we used
the symbol $\times$) played a crucial role in the
proof of background independence. In the same way as this equation
holds strictly for the case of polyhedra, our present equation
\pcomk\ holds strictly for minimal area sections. For carrying out the
proof of background independence we only need relations up to a
choice of local coordinates at the special puncture, since a dimension
(0,0) primary state $\ket{\wh\O_\mu} \equiv \ket{c\bar c \O_\mu}$
will be inserted there.

\chapter{Representation in $\H$ and Quantum Master Action}

In the present section we begin by discussing the construction of
the volume element on $\H$ necessary to define the BV delta operator
$\Delta$.  We then give representation formulae on $\H$ for the geometrical
operations $\p, \Delta\,, \{\, ,\, \},$ and $\K$ introduced in the
previous section. Here we see why the moduli space operators
$\Delta$ and $\{\, ,\, \},$ correspond to the delta and antibracket
of BV quantization. The operator $\K$ turns out to be related
to the action of covariant derivatives $D_\mu (\,\wh\Gamma\,)$, with
$\wh\Gamma$ the canonical connection of
Refs.[\kugozwiebach,\rangaconnection,\rangasonodazw].
We then review, in this new language, the geometrical recursion relations
of closed string field theory, and the basics of the construction
of the quantum closed string  master action. While presenting the
material in this (and later) sections, we shall assume
some familiarity with
Ref.[\senzwiebach], whose notation we follow.

\section{Anti-bracket, Volume Element and Delta Operator}

In Ref.[\senzwiebach] we have described in detail the symplectic form $\omega$
on the state space $\H$ relevant to the formulation of closed string field
theory.
This space is the subspace of the complete state space
$\HH$, and is spanned by the states $\ket{\Phi_i}$ annihilated
by $b_0^-$ and $L_0^-$.
The string field is written as $\ket{\Psi}=\sum_i\ket{\Phi_i}\,\psi^i$, where
the target space fields $\psi^i$ play the role of coordinates in $\H$.
The symplectic form $\bra{\omega}$, and its inverse $\ket{\s}$,
in $\H$ are given by
$$\eqalign{
\bra{\omega_{12}}=\bra{R'_{12}} c_0^{-(2)} &\equiv
\,-\, ~_1\bra{\Phi^i}\,\,\omega_{ij}(x)\,\, ~_2\bra{\Phi^j},\cr
\ket{\s_{12}} = b_0^{-(1)} \ket{R'_{12}} &\equiv \,\,
\ket{\Phi_i}_1\,\,(-)^{j+1}\omega^{ij}(x)\,\,\,\ket{\Phi_j}_2 \,.\cr}
\eqn\emmten$$
Given two functions
$A$ and $B$ in $\H$ one can show that
$$\bigl\{ A \, , B \, \bigr\}\equiv {\partial_r A \over \partial \psi^i}
\omega^{ij} {\partial_lB\over \partial \psi^j} \,=\, (-)^{B+1}
{\p \,A \over \p \,\ket{\Psi}}\, {\p \,B \over \p \,\ket{\Psi}}\, \ket{\s}\, ,
\eqn\xantibr$$
where the sewing ket $\ket{\s}$ is gluing the two state spaces left open
by the differentiation with respect to the string field. There is no need
to specify left or right
derivatives because the string field is even. The kinetic term
of closed string field theory uses the symplectic form. It is given by
$$S_{0,2} = \half\, \bra{\omega_{12}} Q^{(2)} \ket{\Psi}_{1}
\ket{\Psi}_{2} \, . \eqn\dfkterm$$

We must now describe the volume element and the construction of
the BV delta operator $\Delta$, relevant to the formulation of
quantum closed string field theory. In general, in a BV theory,
the volume element $d\mu$ is written as
$$d\mu = \rho (\psi) \prod_i d\psi^i, \eqn\volelement$$
where the product runs over all values of $i$. The volume element is not
a differential form \foot{The volume element acts on a set of vectors, if
we transform this set by a linear transformation $A$, the volume element
transforms with a factor $\hbox{sdet} A$. Such transformation law cannot
be achieved with a linear differential form.}
and does not arise from the symplectic form $\omega$
(since $\omega\wedge\omega =0$).
It is independent data that must be specified.
In the same sense as $\sqrt{g} dx\wedge\cdots dx$, in Riemannian
geometry, it is a coordinate invariant object.
Therefore, an explicit computation of $\rho(\psi)$ requires
a choice of local coordinates.
Associated to the volume element $d\mu$,
the delta operator $\Delta_{d\mu}$ is defined by
$$\Delta_{d\mu}\, A\, \equiv \half \,\hbox{div}_\rho V_A \equiv
\, {1\over 2\rho} (-)^i {\p_l\over \p\psi^i}\,\Bigl(\, \rho\, \omega^{ij}
{\p_l\over \p\psi^j}\, A \, \Bigr) , \eqn\deltaop$$
where $A$ is a scalar function, and $\Delta_{d\mu}\, A$ is the scalar
obtained by taking (one-half of) the divergence of the hamiltonian
vector field $V_A$. There is an important consistency condition,
one must choose $\rho(\psi)$ such that $\Delta_{d\mu}^2 =0$ [\schwarz].
As usual, $\p_l$ and $\p_r$ denote left and right derivatives respectively.

For the case of closed string field theory, in the chosen basis for $\H_x$,
we take $\rho = \rho(x)$, which is field independent, but can, in general,
depend on the coordinate $x$ labeling the particular CFT around which
string field theory is being formulated,
$$d\mu = \rho(x)\prod_i d\psi^i_x, \quad\to\quad
\Delta_{d\mu}\, A\,=
\, \half\, (-)^i {\p_l\over \p\psi^i_x}\,\Bigl(\, \omega^{ij}
{\p_l\over \p\psi^j_x}\, A \, \Bigr) =\, \half\, \hbox{str}\Bigl[\,
{\p_l\over \p\psi^k_x}\,\Bigl(\, \omega^{ij}
{\p_l\over \p\psi^j_x}\, A \, \Bigr)\Bigr]\,.\eqn\selement$$
where str$(\wt A_k^{~i})\equiv (-1)^i \wt A_i^{~i}$.
Since the $\omega^{ij}$'s are constants with statistics
$(-)^{(i+j+1)}$, and satisfy $\omega^{ji}=-(-)^{(i+1)(j+1)}\omega^{ij}$,
one readily verifies that
$\Delta_{d\mu}$ is nilpotent. While the field independent $\rho(x)$ factor
in the measure drops out of the delta operator, it will be relevant for
us since we will be comparing theories at different points $x$, and there
is no a priori reason why this scale factor should be a constant on theory
space. With the same methods used in
Ref.[\senzwiebach] to prove \xantibr\ we can verify that
for any function $U$ on $\H$,
$$\Delta \, U = \,\half\, (-)^{U+1} \Bigl(\, {\p\over\p\ket{\Psi}}
{\p\over\p\ket{\Psi}}\, U \Bigr) \, \ket{\s}\, .\eqn\takedelta$$

In the BV formalism observables are obtained by integration,
with suitable measures, over lagrangian submanifolds (of the supermanifold
$M$ of field/antifield configurations). A lagrangian submanifold $L$
is defined by the condition that at any point $p\in L$, for any two
tangent vectors $e_i, e_j \in T_pL$, we have $\omega (e_i , e_j ) = 0$.
The volume element $d\mu$ in $M$ then induces a volume element $d\lambda$
on $L$ as follows [\schwarz]. Let $p\in L$, and $(e_1,\ldots,e_n)$
be a basis of $T_pL$. One then defines
$$d\lambda (e_1,\cdots,e_n)\equiv \, [ d\mu
(e_1,\cdots,e_n,f^1,\cdots,f^n)\, ]^{1/2}\ ,\eqn\LAMBDA$$
where the vectors $f$ are any set of vectors satisfying
$\omega(e_i,f^j)=\delta_i^j$. This condition fixes the vectors
$f^j$ up to the transformation $f^j \to f^j + C^{ji} e_i$. The right hand
side of \LAMBDA, however, is invariant under this transformation, since
it corresponds to a transformation of the complete basis
$(\{ e_i\} ; \{ f^j\})$ by a matrix of unit superdeterminant.

\section{Representation on $\H$.}

\subsection{Basic Representation Formulae.}
In conformal field theory surfaces are represented by surface states,
and spaces of surfaces can therefore be represented by integrals of surface
states. The operators $\p$, $\Delta$ and $\{ \, , \, \}$,
discussed above, act on
spaces of surfaces. They can, as a consequence, be
represented by operators acting on integrals of surface states.
We are familiar with the correspondence $\p \leftrightarrow Q$
leading to
$$ \int_{\A_{g,n}} \bra{\Omega^{(k)g,n}} \sum_{i=1}^n Q^{(i)}
=(-1)^k \int_{\p\A_{g,n}} \bra{\Omega^{(k-1)g,n}} .\eqn\eketbrst$$
Here $\bra{\Omega^{(k)g,n}_\Sigma}$ denotes  (upon contraction
with $n$ arbitraty states in $\H$),
a $(6g\hskip-2pt +\hskip-2pt 2n\hskip-2pt -6\hskip-2pt +k)$-form
on the moduli space $\wh\P_{g,n}$.
We generally omit from the form the label $\Sigma$
corresponding to the surface. These forms are
explicitly given by [\zwiebachlong,\kugosuehiro\ ] ~\foot{The normalization
factor includes an extra minus sign that went unnoticed in
Ref.[\zwiebachlong].}
$$\bra{{\Omega_{}^{}}^{(k)g,n}}( V_1,\cdots , V_{6g+2n -6+k} )
= (-2\pi i)^{(3-n-3g)}\bra{\Sigma}\,{\bf b}({\bf v}_1)\cdots
{\bf b}({\bf v}_{6g\hskip-1pt +2n-6+k}).
\eqn\cdefform$$
The Schiffer vector ${\bf v}_r= (v_r^{(1)}(z),\cdots v_r^{(n)}(z))$
creates the deformation of the surface $\Sigma$ specified by the tangent
$V_r$, and the antighost insertions are given by
($\ointop dz/z=\ointop d\bar z/\bar z=2\pi i$)
$${\bf b}({\bf v}) = \sum_{i=1}^n \biggl(
\oint b^{(i)}(z_i) v^{(i)}(z_i) {dz_i\over 2\pi i}
+\oint \overline b^{(i)}( \overline z_i)  \overline v^{(i)}
(\overline z_i) {d\overline z_i\over 2\pi i} \biggr).\eqn\kjhkjh$$

The sewing ket $\ket{\s}$ implements, at the level of states, the action
of the bracket $\{ \, , \, \}'$, or $\Delta$ at the level of surfaces.
If the two punctures to be sewn are on different surfaces we have
(see refs.[\zwiebachlong, \senzwiebach] for details):
$$\Big(\hskip-6pt\int_{\A_{g_1, n_1}}
\hskip-6pt\bra{\Omega^{(k_1)g_1,n_1}}
\int_{\A_{g_2, n_2}}
\hskip-6pt\bra{\Omega^{(k_2)g_2,n_2}}\Big) \ket{\s}
 \,=\,\,(-)^{k_2}\hskip-16pt\int_{\{ \A_{g_1, n_1}\, , \,  \A_{g_2, n_2}\}'}
\hskip-13pt\bra{\Omega^{(k_1+k_2-1)g_1+g_2,n_1+n_2-2}} \, ,
\eqn\esewingtimes$$
where the sewing ket contracts with the state spaces corresponding
to the punctures being sewn by the bracket.
The factor $(-)^{k_2}$ arises from the necessity of moving the tangent vector
associated with the sewing angle through the tangent vectors of
$\A_{g_2,n_2}$. When the punctures to be sewn are on the same surface
we have
$${1\over 2} \int_{\A_{g,n}}\, {}_{1\cdots n}\bra{\Omega^{(k)g,n}} \s_{12}
\rangle =\,(-)^k \int_{\Delta\A_{g,n}} \, {}_{3\cdots n}
\bra{\Omega^{(k-1)g+1,n-2}}\, . \eqn\esewingdelta$$
For the case of $A_{0,3}$, a space whose elements
are three punctured spheres, the definition of $\Delta$ implies that
the space $\Delta \A_{0,3}$ is formally ${1\over 2}$ of a space
$\A'_{1,1}$ in $\wh\P_{1,1}$.
The integral on the right
hand side of the equation should be interpreted as $\half \int_{\A'_{1,1}}$.
This factor of $\half$ multiplying an integral over a space of surfaces
is natural on account of the observation of Ref.[\polchinski]
that a punctured torus has a $Z_2$ group of diffeomorphism (generated by
$z\to -z$), and hence the expression for the one loop partition function
(or one point function) should automatically carry a factor of ${1\over
2}$. This factor of ${1\over 2}$ will be relevant in our
analysis of \S6.

\subsection{Representation of the BV algebra.}
In order to make manifest why our operations on moduli spaces
become the standard BV operations on $\H$, we consider a canonical
procedure that, given a subspace of surfaces, associates to this
subspace a functions on $\H$. Let $\A_{g,n}^{(k)}$ denote a
(basic) subspace of $\wh\P_{g,n}$ of real dimensionality $(6g-6+2n+k)$.
We define the associated function $f(\A_{g,n}^{(k)})$ as
$$f(\A_{g,n}^{(k)}) \,\equiv {1\over n!}\,
\int_{\A_{g,n}^{(k)}}\bra{\Omega^{(k)g,n}}\Psi\rangle_1\cdots
\ket{\Psi}_n\,\, , \quad n\geq 1\, .\eqn\fgeomty$$
where, for convenience of writing, we do not include $\ket{\Psi}$ as an
argument of $f$.
This is clearly a very natural operation, we are simply integrating the
canonical forms on moduli space over the subspace of surfaces.
This definition
is complete for spaces of surfaces of genus zero with three or more
punctures, and for spaces of surfaces of higher genus with one or more
punctures.
For $g\ge 2, n=0$, we define,
$$ f(\A^{(k)}_{g,0}) \equiv \int_{\A^{(k)}_{g,0}} \Omega_\Sigma^{(k)g,0}\,,
\eqn\edefffornzero$$
where,
$$ \Omega^{(k)g,0}_\Sigma \equiv \, (-2\pi i)
\bra{\Omega^{(k-2)g,1}_{\wh\Sigma}}
0\rangle \,.\eqn\edefomegagzero$$
Here $\Sigma$ is a genus $g$ surface without punctures, and
$\Omega^{g,0}_\Sigma$ is a $(6g-6+k)$ form on the tangent space to
$\M_{g,0}$ at $\Sigma$.
To construct this form, we must first construct a surface
$\wh\Sigma\in\wh\P_{g,1}$, by introducing a puncture (with a local coordinate)
on $\Sigma$. Then, as indicated in the above equation, we use a form
on $T_{\wh\Sigma}\wh\P_{g,1}$ to define the form $\Omega^{(k)g,0}_\Sigma$.
More precisely, given $(6g-6+k)$ tangents
$(\vec{V}_1,\cdots \vec{V}_{6g-6+k}) \in T_\Sigma \M_{g,0}$, we must choose
$(6g-6+k)$ tangents
$(\wh{V}_1,\cdots \wh{V}_{6g-6+k}) \in T_{\wh\Sigma}\wh\P_{g,1}$, which project
down to the original tangents as we forget about the extra puncture. The
definition in \edefomegagzero\ then reads
$$\Omega^{(k)g,0}_\Sigma (\vec{V}_1,\cdots \vec{V}_{6g-6})
\equiv \, (-2\pi i)\bra{\Omega^{(k-2)g,1}_{\wh\Sigma}}0\rangle
(\wh{V}_1,\cdots \wh{V}_{6g-6})\eqn\moreexpli$$
It is known that this definition
is independent of the choice of surface $\wh\Sigma$ projecting to $\Sigma$,
and of the tangents in $T_{\wh\Sigma}\wh\P_{g,1}$ projecting to the given
tangents in $T_\Sigma \M_{g,0}$
[\alvarez]. Also, since $\bra{\Omega^{(k-2)g,1}_{\wh\Sigma}}$ carries
ghost number $6-k$, it is clear from Eqn.\edefomegagzero\ that
$\Omega^{(k)g,0}$ vanishes unless $k=0$. This gives,
$$ f(\A^{(k)}_{g,0})=0 \quad \hbox{for} \, \, k\ne 0\, . \eqn\fvanishing $$
There is no completely natural definition of $f$ for $g=1, n=0$, but
we shall define an analog of the function $f$ in \S6.

The definition of $f$ on the complete complex $\wh\P$ of surfaces follows from
$$f\,\Big( \sum_i\,\,a_i \A_{g_i,n_i}^{(k_i)} \Big) =\sum_i\,\, a_i\,
f\,(\A_{g_i,n_i}^{(k_i)})\,.\eqn\repprod$$
We now claim that for $\A, \B\in\wh \P$, the following representation
identities hold:
$$\eqalign{
f\bigl( \Delta \A ) &= -\Delta  f(\A) \, \cr
f \bigl( \{ \A , \B \} \bigr) &= - \{\, f(\A) , f(\B) \}\, .\cr}\eqn\hreh$$
On the left hand side of the above equations, the delta operator and the
antibracket act on the spaces of surfaces, on the right they act on the
functions in $\H$. These equations justify our claim that the bracket
and delta operators on moduli space give rise, by representation on
functions in $\H$ to the BV antibracket and delta operators.
More precisely, these define a homomorphism of the Riemann surface
algebra of $\Delta$ and the antibracket, to the corresponding algebra of
string functionals.
Eqns.\hreh\ are first proven for basic spaces, and then are easily extended
to generalized spaces. Let us prove the second of these equations as
an illustration. Making use of \xantibr\ we find
$$\eqalign{
\bigl\{ f(\A_{g_1,n_1}^{(k_1)})\, ,f(\A_{g_2,n_2}^{(k_2)})\, \bigr\} &=
(-)^{k_2+1} \,
{\p f(\A_{g_1,n_1}^{(k_1)})\over \p \ket{\Psi} }\,
{\p f(\A_{g_2,n_2}^{(k_2)})\over \p \ket{\Psi} }\, \ket{\s}\,,\cr
&= (-)^{k_2+1} \int_{\A_{g_1,n_1}^{(k_1)}}  \bra{\Omega^{(k_1)g_1,n_1}}
\,{\ket{\Psi}^{n_1-1} \over (n_1-1)! }
\int_{\A_{g_2,n_2}^{(k_2)}} \bra{\Omega^{(k_2)g_2,n_2}}\,
\,{\ket{\Psi}^{n_2-1} \over (n_2-1)! }\ket{\s}\,,\cr
&= \,\,-\hskip-15pt\int_{ \{ \,\A_{g_1,n_1}^{(k_1)}\,,
\A_{g_2,n_2}^{(k_2)}\}' }
\hskip-15pt\bra{\Omega^{(k_1+k_2 -1)g_1+g_2,n_1+n_2 -2} }
{\ket{\Psi}^{n_1+n_2-2} \over (n_1-1)!(n_2-1)! }\,,\cr}
\eqn\drvantb$$
where in the last step we made use of \esewingtimes.  {}From the relation
between the primed antibracket and the standard antibracket (Eqn.\rbracket)
we now obtain
$$\eqalign{
\bigl\{ f(\A_{g_1,n_1}^{(k_1)})\, ,f(\A_{g_2,n_2}^{(k_2)})\, \bigr\} &=
\,\,\,- \hskip-15pt\int_{ \{ \,\A_{g_1,n_1}^{(k_1)}\,, \A_{g_2,n_2}^{(k_2)}\}}
\hskip-15pt\bra{\Omega^{(k_1+k_2 -1)g_1+g_2,n_1+n_2 -2} }
{\ket{\Psi}^{n_1+n_2-2} \over (n_1+n_2-2)! }\,,\cr
&= -f\bigl( \{ \,\A_{g_1,n_1}^{(k_1)}\,,  \A_{g_2,n_2}^{(k_2)}\}\bigr) \,,\cr}
\eqn\drvantb$$
as we wanted to show. The first identity in Eqn.\hreh\ follows in a
similar manner through the use of Eqn.\esewingdelta.
One more identity will be relevant to our analysis. From Eqns.\xantibr\
and \eketbrst\ it follows that
$$
\{ \,S_{0,2} , f (\A )\} = \, - \,
f\bigl( \p \A\bigr)
\, ,\eqn\reppd$$
where $S_{0,2}$ is the kinetic term of closed string field theory,
as given in Eqn.\dfkterm. Thus the operation of taking boundary is represented
in $\H$ by a canonical transformation induced by the kinetic term of
closed string field theory.

\subsection{Representation identities for $\K$.}
A specific  connection $\wh\Gamma$ in the CFT theory space was introduced
in Refs.[\kugozwiebach,\rangaconnection,\rangasonodazw]
and further explored in
Ref.[\senzwiebach]. It is defined by the relation
$$D_\mu (\,\widehat\Gamma\,) \bra{\,\Sigma\, } = -{1\over\pi} \hskip-6pt
\int_{\Sigma - \cup_i D_i^{(1)}}\hskip-6pt d^2 z~\bra{\,\Sigma;z\,}
\O_\mu\rangle\, .\eqn\xen$$
It follows from this relation that for spaces of surfaces and string field
forms $\bra{\Omega}$ we have
$$ D_\mu(\,\wh\Gamma\,) \int_{\A_{g,n}} \, \bra{\Omega^{(k)g,n}}
= \hskip-3pt\int_{\K\A_{g,n}}
\hskip-5pt\bra{\Omega^{(k)g,n+1}} \wh\O_\mu\rangle_{n+1}\,,\eqn\edmugomega$$
where the $(n+1)$-th puncture in the right hand side
is the special puncture of $\K\A_{g,n}$.
Equation \edmugomega\ is verified
using the fact that the connection $\wh\Gamma$ does not act on antighost
insertions, the relation between string forms and surface states
(Eqn.\cdefform), and the relation
$dx\wedge dy \, b (\p/\p x) b(\p/\p y) \ket{\wh\O_\mu}
= 2idx\wedge dy\,\ket{\O_\mu}$, which requires the use of the definition given
in Eqn.\kjhkjh.

We define now the map $f_\mu$ which acts on spaces of surfaces of two
types; surfaces with one special puncture, or surfaces
with no special puncture.
Acting on such spaces it gives us a function in $\H$. We take
$$f_\mu(\A_{g,n+1}^{(k)}) \,\equiv
{1\over n!} \int_{\A_{g,n+1}^{(k)}} \bra{\Omega^{(k)g,n+1}\,}
\Psi\rangle_1\cdots\ket{\Psi}_{n}\ket{\wh\O_\mu}\, .\eqn\fgeomty$$
For spaces of surfaces with a special puncture, the state $\ket{\wh\O_\mu}$
is inserted at that puncture.
For spaces with no special puncture, $\ket{\wh\O_\mu}$
can be inserted at any puncture (since anyway all our spaces are symmetric).
It follows from this definition and Eqn.\edmugomega\ that
$$ \hbox{D}_\mu (\wh\Gamma \, )  \, f ( \A ) = f_\mu \, \bigl( \K\A
\bigr)\,. \eqn\rprcder$$
where the operation $\hbox{D}_\mu(\wh\Gamma)$, acting on a function of
$\Psi$, has been defined in Eqn.(3.5) of Ref.[\senzwiebach].
Here $\A$ is a space of surfaces without a special puncture, and
$\K\A$ is a space with one special puncture. Whereas  the validity of
Eqn.\rprcder\ for $\A\subset\wh\P_{g,n}$ with $n\ge 1$,
is an immediate consequence of Eqn.\edmugomega,
the extension to the case $n=0$ involves more work.
It has been shown in appendix A that,
$$ \p_\mu \int_{\A_{g,0}} \Omega_\Sigma^{(g,0)} = \int_{\K\A_{g,0}}
\bra{\Omega^{(0)g,1}} \wh\O_\mu\rangle \eqn\evacuumthreex $$
where $\Omega_\Sigma^{(g,0)}\equiv \Omega_\Sigma^{(0)g,0}$
is the top form in $\M_{g,0}$ whose
definition was given in Eqn.\edefomegagzero.
Since acting
on $\ket{\Psi}$ independent terms $\hbox{D}_\mu\equiv \p_\mu$,
Eqn.\evacuumthreex\ implies that Eqn.\rprcder\ holds also for $\A\subset
\wh\P_{g,0}(\equiv\M_{g,0})$, and hence for any $\A\in \wh\P$.

It is also straigthforward to verify that equations analogous to \hreh\
hold for the function $f_\mu$.
We find that
$$\eqalign{
f_\mu \bigl( \Delta \A ) &= -\Delta  f_\mu(\A) \, \cr
f_\mu \bigl( \{ \A , \underline{\B} \} \bigr) &=
- \{\, f(\A) , f_\mu(\B) \}\, ,\cr}\eqn\hrehn$$
where in the second equation the space $\B$ on the left hand side was
underlined to denote the fact that this space has a special puncture.

There is an odd hamiltonian function on $\H$ that by a canonical
transformation inserts a special puncture. We define
$${\bf U}_{\mu (0,2)}\equiv \bra{\omega_{12}\,}\wh\O_\mu\rangle_1\ket{\Psi}_2
\,\, , \eqn\shftpsi$$
and one can readily prove that for a space $\A$ without a special puncture
$$f_\mu (\A) = \,\{\, f(\A)\, , \,{\bf U}_{\mu
(0,2)}\,\}\,\,.\eqn\createp$$
Two additional equations are obtained easily. We have
$$\{ \,S_{0,2} \,,\, f_\mu (\A)\}= \, - \, f_\mu(\p\A)\,, \eqn\bounfmu$$
which is the exact analog of Eqn.\reppd, and
$$ \{ \, S_{0,2} \,,\, {\bf U}_{\mu(0,2)} \} = 0\,,\eqn\vnshcon$$
which follows as a result of $Q \ket{\wh\O_\mu} =0$.
Finally, we can show that for a space of surfaces $\A$ with one
special puncture (and any number of ordinary ones)
$$ \A \approx 0\quad\to\quad f_\mu (\A ) = 0\,. \eqn\basicint$$
This is obtained as follows. By definition $\A \approx 0$ means
$\pi_f (\A)=0$, where $\pi_f$ was defined above Eqn.\distrk.
Since $L_{n\geq0}$ and $b_{n\geq 0}$, together with their anti-holomorphic
counterparts,
annihilate $\ket{\wh\O_\mu}$, it
follows that the form  $\bra{\Omega}\Psi\rangle\cdots\ket{\wh\O_\mu}$ on $\A$
which enters in the definition of $f_\mu$, descends to a well defined form in
$\pi_f (\A)$. Since $\pi_f$ must be locally a diffeomorphism (otherwise
$\A$ must have some tangent vector
representing change of coordinates
around the special puncture,  and the corresponding
antighost insertion will give $b_n\ket{\wh\O_\mu}$ for $n\ge 0$,
making $f_\mu$ vanish trivially) the integral
over $\A$ defining $f_\mu$ can be written as an integral over $\pi_f(\A)$,
and therefore vanishes since $\pi_f(\A) =0$.
This establishes \basicint.

\section{Master Action of Closed String Field Theory}

The complete quantum master action for closed string field theory is given by
the sum of the kinetic term plus geometrical terms corresponding to integrals
over string vertices. Using the map $f$ introduced earlier we simply have
$$S= S_{0,2} + f (\V) + \hbar S_{1,0}\,,\eqn\fullaction$$
where $\V$ and $S_{0,2}$ were given in Eqns.\dfnvsp\ and \dfkterm.
Here we have included the vacuum vertices for $g\ge 2$ in $\V$, but have
written down the contribution of the $g=1, n=0$ term separately.
The kinetic term of the theory is easily
shown to satisfy [\zwiebachlong]
$$\{ S_{0,2} \,,\, S_{0,2}\,\} = 0 \,,\quad\hbox{and},\quad
 \Delta S_{0,2} =0.\eqn\bpropkin$$
Our work proving representation identities allows us to give
a straightforward proof that the action satisfies the master equation. Using
Eqns.\fullaction\ and \bpropkin\  we find
$$\eqalign{
\hbar\Delta S+ \half\{S,S\}& = \hbar\Delta f (\V) + \{\,S_{0,2}\,,\,f(\V)\,\}
+ \half \{ f(\V) \,,\,f(\V)\,\}\,,\cr
&= -\hbar f\, (\Delta \V ) - f\,( \p\V ) -\half f\,(\,\{ \V , \V \}\, ) \,,\cr
&= -f\, (\hbar\Delta \V + \p\V  +\half\,\{ \V , \V \}\, ) =0\,,\cr}
\eqn\proveme$$
where we used Eqns.\hreh\ and \reppd, and the last expression vanishes
by virtue of the geometrical recursion relations \recrelnew.\foot{Note
that $\Delta\V_{g,0}$ vanish identically, and $\V_{g,0}$
does not contribute to $\{\V, \V\}$. Furthermore, $\p\V_{g,0}$ is of
dimension $(6g-7)$, and hence $f(\p\V_{g,0})$ vanishes by Eqn.\fvanishing.
This shows that the vacuum graphs are
irrelevant to the verification of the master equation.}

Our proof of background independence will require the computation
of the covariant derivative $\hbox{D}_\mu (\wh\Gamma\,) S$. We first compute
the derivative of the kinetic term
$$\eqalign{
\hbox{D}_\mu (\,\wh\Gamma\, ) S_{0,2} &= \half\, \Bigl(
 D_\mu (\wh\Gamma ) \bra{\omega_{12}} Q^{(2)} \Bigr)
\ket{\Psi}_1\ket{\Psi}_2\, ,\cr
&=\half\bra{V'^{(0,3)}_{123}}\wh\O_\mu\rangle_3\,\ket{\Psi}_1\ket{\Psi}_2\,,\cr
&= f_\mu (\V'_{0,3})\,,\cr}\eqn\takederk$$
where we made use of Eqn.(3.23) of Ref.[\senzwiebach], the definition
of $f_\mu$ in Eqn.\fgeomty, and the definition
$\bra{V^{\prime(0,3)}}\equiv
\bra{\Omega^{(0)0,3}}_{\V'_{0,3}}$. The complete result for the action now
follows from this result for the kinetic term and Eqn.\rprcder. We have
$$\hbox{D}_\mu (\,\wh\Gamma\, ) S =
f_\mu (\V'_{0,3}) +\hbox{D}_\mu (\,\wh\Gamma\, ) f (\V) +\hbar\p_\mu S_{1,0}
= f_\mu (
\V'_{0,3}+ \K\V )+ \hbar\p_\mu S_{1,0}\,.\eqn\fcovder$$
where $\p_\mu\equiv {\p\over \p x^\mu}$. Since $S_{1,0}$ is
string field independent, we have replaced $\hbox{D}_\mu(\wh\Gamma) S_{1,0}$
by $\p_\mu S_{1,0}$.

\chapter{Setup for Quantum Background Independence}

The objective of the present section is to give the conditions for
quantum background independence of closed string field theory. The
physical and geometrical motivation for the proposed conditions will
be given, and we will derive their explicit form
for the case of nearby backgrounds.
The main point that we make here is that the action weighted measure
$d\mu_S \equiv  d\mu e^{2S/\hbar}$ is the background independent
object. As studied recently in Ref.[\senzwiebachgauge] this measure
encodes all of the physics of the theory, and quantum gauge transformations
leave it invariant. For a discussion of $d\mu_S$ from a somewhat different
perspective see Ref.[\schwarznew].

\section{The Conditions of Quantum Background Independence}

The question of quantum background independence of string
field theory is formulated as follows. Let $x$ and $y$ be two
conformal field theories, and let $\H_x$ and $\H_y$ be their
respective state spaces, equipped with symplectic structures
and measures $(\omega_x , d\mu_x)$ and $(\omega_y , d\mu_y)$
respectively. Moreover, we have master actions
$S_x : \H_x \to R$
and $S_y : \H_y \to R$ satisfying the respective master equations.
We demand the
existence of a diffeomorphism
$$F_{y,x} :\H_x\to \H_y\, ,\eqn\diffeo$$
between the corresponding spaces $\H_x$ and $\H_y$
such that
$$\eqalignno{\omega_x &={F_{y,x}}^*\,\omega_y\,\, ,&\eqname\one\cr
\quad d\mu_x\, e^{2S_x/\hbar} &={F_{y,x}}^*\,
\bigl( d\mu_y\,e^{2S_y/\hbar}\bigr)\,\, ,&\eqname\two\cr}$$
with $F_{y,x}^{~~~*}$ denoting the pullback performed using the diffeomorphism
$F_{y,x}$. The first of these equations implies that the diffeomorphism
preserves the antibracket, and the second equation indicates that the
action-weighted volume elements $(d\mu_S)_x$ and $(d\mu_S)_y$
must be taken into each other.
The question of quantum background independence is simply the question whether
such symplectic diffeomorphism exists.

\subsection{Geometrical Interpretation}
The above conditions for background independence might seem a bit
strange at first sight. A Batalin-Vilkovisky (BV) manifold is defined by
$(M,\omega,d\mu)$ where
$M$ is a supermanifold, $\omega$ is a symplectic form, and $d\mu$ is
a volume element. The volume element must lead to a nilpotent
$\Delta_{d\mu}$. The space $\H$ relevant for string field theory
is such a BV manifold; the symplectic structure
was given in [\senzwiebach], and the volume element
$d\mu = \rho \prod_i d\psi^i$ leads to a nilpotent $\Delta_{d\mu}$.
One might be tempted to assume
that a proof of background independence at the quantum level would
have meant the existence of a diffeomorphism
mapping the BV manifolds
$(\H_x,\omega_x,d\mu_x)$ and $(\H_y,\omega_y, d\mu_y)$
into each other, and at the
same time taking one master action into the other.
This is not the case. One cannot find a symplectic map that
preserves the volume element and the action separately. In fact, the
existence of such map is
not necessary physically, as will be explained below. There is, however,
a simple sense in which quantum background independence is
strictly a statement that two spaces are identical BV manifolds.

Assume we have a volume element $d\mu$ leading to a nilpotent $\Delta_{d\mu}$.
Consider now the measure $d\mu_S =d\mu\,e^{2S/\hbar}$. The associated delta
operator is given by
$$\Delta_{d\mu_S} = \Delta_{d\mu} +
{1\over \hbar}\,\{ S\, ,\,\, \cdot \,\, \} \eqn\three$$
and then, one can show (following Schwarz[\schwarz])
$$\Delta_{d\mu_S}^2  = {1\over \hbar^2}\,
\Bigl\{ \, \hbar\,\Delta_{d\mu} \,S\,+ {1\over 2}\{\, S\,,\,S\,\}\,,\,\,
\cdot \,\, \Bigr\}\,\,, \eqn\four$$
which shows that $\Delta_{d\mu_S}^2$ is a linear operator,
in fact, a hamiltonian vector field.
This equation also shows that $d\mu_S$ is a consistent
measure, leading to a nilpotent $\Delta_{d\mu_S}$,
if $d\mu$ is consistent, and $S$ satisfies the quantum master equation.
This implies that given a BV manifold $(\H,\omega,d\mu)$, we can define
another BV manifold $(\H,\omega,d\mu_S)$.
Quantum background independence is the statement that
$(\H_x,\omega_x,[d\mu_S]_x)$, and $(\H_y,\omega_y,[d\mu_S]_y)$
are the same BV manifold.

\subsection{Physical Interpretation}
The conditions of background independence should imply
that the two string field theories, formulated around different
conformal backgrounds, are just two different descriptions of
the same underlying theory. More precisely,
the conditions of background independence imply a {\it formal} equivalence
between the theories formulated using the different backgrounds.
The emphasis on the word formal is necessary, the physics around
different backgrounds can, in general, be different. As will be discussed
below, this is due to the constant shift in the diffeomorphism
relating the two theories.

Let us see why the conditions of background independence imply
the formal equivalence of the two theories.
The observables in a theory are defined by
$$\vev{A} \equiv \int_L \, d\lambda e^{S/\hbar} \, A\, ,\eqn\five$$
where $L$ denotes a Lagrangian submanifold, and $A$, in order
to be a physical operator, must be a function
of fields/antifields satisfying
$ \hbar\, \Delta_{d\mu} A + \{ S , A\} = 0$.
This condition, with the help of Eqn.\three\ reads
$\Delta_{d\mu_S}\, A = 0$.
This is important, as it shows that knowledge of the measure
$d\mu_S$ suffices to define physical operators.

Consider now the case when we have two string field theories formulated
around conformal theories $x$ and $y$. Let $A_y$, satisfying
$\Delta_{d\mu_{S_y}}\, A_y=0$, be an observable of the string theory
at $y$.  Since condition \two\ makes
$\Delta_{d\mu_S}$ a scalar operator,
$A_x \equiv F_{y,x}^{~~~*}A_y$ satisfies
$\Delta_{d\mu_{S_x}}\, A_x = 0$, and is therefore a physical
operator of the theory at $x$. Since $F_{y,x}^{~~~*}$ is symplectic,
the preimage
$L_x \equiv F_{y,x}^{~~~*}L_y$ of $L_y$
is a lagrangian submanifold in $\H_x$. Finally, it follows from
\two\ and \LAMBDA\ that
$$\quad d\lambda_x\, e^{S_x/\hbar} ={F_{y,x}}^*\,
\bigl( d\lambda_y\,e^{S_y/\hbar}\bigr)\,\, .\eqn\nine$$
All this put together implies the formal equality of observables in
the two string theories
$$\vev{A_x}\equiv \int_{L_x} \, d\lambda_x e^{S_x/\hbar} \, A_x\,=\,
\hskip-6pt\int_{F_{y,x}^*L_y} \hskip-5pt{F_{y,x}}^*\,
\bigl( d\lambda_y\,e^{S_y/\hbar} A_y\, \bigr) =
\int_{L_y} \, d\lambda_y e^{S_y/\hbar} \, A_y \,=\vev{A_y}\,.\eqn\ten$$
This proves that Eqns.\one\ and \two\ guarantee the formal
equivalence of the theories formulated around different backgrounds.

Does the existence of a solution of Eqns.\one\ and \two\ imply that string
theories formulated around different points in the CFT moduli space
correspond to the same quantum theory? To answer this question, we begin with
the observation that in computing correlation functions of observables
using the path integral formalism, we need not only the action weighted
measure, but also the asymptotic
boundary conditions on the fields. More than one possible boundary conditions
may be allowed in general, giving rise to different consistent quantum
theories with different sets of correlation functions.\foot{Examples of
such theories already exist in the context of $c\le 1$ string
theories[\senmatrix,\seibergshenker].} These different
correlation functions may be interpreted as expectation values of
observables calculated in different ground states. Thus the question of
uniqueness of the correlation functions is related to the uniqueness of
possible allowed
asymptotic boundary conditions on the fields, or, equivalently, uniqueness
of the ground state of the theory. If there are different allowed ground
states of the theory, we may refer to them as different phases of the
same theory.

Normally, when we work with an action $S_x(\ket{\Psi_x})$, it is
implicitly assumed that we use
the boundary condition $\ket{\Psi_x}\to 0$ asymptotically. Existence of an
$F_{y,x}$ satisfying Eqns.\one\ and \two\ guarantee that doing a path
integral with measure $d\lambda_x e^{S_x}$ and asymptotic boundary
condition
$\ket{\Psi_x}\to 0$ is equivalent to doing a path integral with measure
$d\lambda_y e^{S_y}$ with asymptotic boundary condition
$\ket{\Psi_y}\to
F_{y,x}\,(\,\ket{\Psi_x}=0\,)$. The result of such a path integration will, in
general, be different from the one where we use the same action weighted
measure, but put the asymptotic boundary condition $\ket{\Psi_y}\to 0$. In this
sense the perturbative correlation functions, computed around two different
CFT backgrounds $x$ and $y$ will be different. This would
seem to imply that we must necessarily have different phases of
string theory associated
with different points $x$ in the CFT moduli space, since for each point
$x$ we have an associated asymptotic boundary condition $\ket{\Psi_x}\to 0$.
This, however, could be an artifact of perturbation theory,
since $\ket{\Psi_x}\to 0$ is an allowed boundary condition only in
perturbation theory. It may turn out that in full quantum string theory,
only a subset of these are allowed asymptotic boundary conditions.
This would be the case, for example, if the degeneracy between the
different ground states at the classical level is lifted by quantum
corrections. In fact, for bosonic string theory, probably none of the
above choices of boundary conditions give rise to a consistent quantum theory
due to the presence of the tachyon. For superstring theories, on the other
hand, each of the above boundary condition seems to give rise to a consistent
quantum theory,
at least to all orders in perturbation theory, if the corresponding tree
level string theory has unbroken $N=1$ space-time supersymmetry.

To summarise, the question we address in our analysis is the background
independence of the action weighted measure $d\mu_S$.
The question of what are the possible allowed asymptotic
boundary conditions, or, equivalently, what are the possible ground states
of the theory, is a dynamical question,
and is beyond the scope of the present analysis.

\section{The Conditions of Local Background Independence}

Let us now find the explicit form that the background independence
conditions take when we consider string theories formulated around
nearby backgrounds. From the expressions for the measures $d\mu_x = \rho(x)
\prod_i d\psi^i_x$, and $d\mu_y = \rho(y)\prod_i d\psi^i_y$,  we readily find
$${F_{y,x}}^*\,
\bigl( d\mu_y\,\bigr) =\, {\rho(y)\over \rho(x)}\,\,
\hbox{sdet} \biggl[ {\p_l\psi_y\over \p\psi_x}\biggr]
\cdot d\mu_x\,.\eqn\transf$$
This result, back into condition \two\ of background independence gives
$$ \exp \Bigl({2S(\psi_x ,x)\over\hbar} \Bigr) =
\exp \Bigl({2S(\psi_y ,y)\over\hbar} \Bigr) \cdot
\, {\rho(y)\over \rho(x)}\,\,\cdot
\hbox{sdet} \biggl[ {\p_l\psi_y\over \p\psi_x}\biggr] \,\,.\eqn\condbi$$
This is an exact equation constraining the map relating theories at
$x$ and $y$.
We now consider, as we did in [\senzwiebach], the infinitesimal
diffeomorphism relating a theory at $x$ to one at $y = x+\delta x$. We write
$$\psi^i_{x+\delta x} = F^i \,(\psi_x , x, x+\delta x) = \psi^i_x
+ \delta x^\mu \, f^i_\mu  (\psi_x , x) + \O (\delta x^2)\, .\eqn\indiff$$
In this case Eqn.\condbi\ expressing the background independence of
the measure $d\mu_S$ reduces to the condition
$${\p S(\psi, x)\over \p x^\mu} +
{\p_r S(\psi_x, x)\over \p \psi^i_x} f^i_\mu  +\half\,\hbar \,\Bigl[
{\p\ln\rho\over \p x^\mu}\, +\,\hbox{str} \Bigl( {\p_l f^i_\mu\over \p\psi^j}
\Bigr)\,\Bigr] \, =\, 0\, .\eqn\getcond$$

In order to give a geometrical formulation for the background independence
condition, it is convenient to define the object $U_\mu^i$ through the relation
$$f^i_\mu \equiv -\wh\Gamma_{\mu j}^{~~i} \,\psi^j \, -\,U^i_\mu\,,\eqn\comd$$
where we have separated out a term proportional to the
connection $\wh\Gamma_\mu$.
Using \comd, we  rewrite \getcond\ in the following form
$$\hbox{D}_\mu(\,\wh\Gamma\,) \, S- \, {\p_r S\over \p \psi^i}
\, U^i_\mu \,+\, \half\,\hbar\,\Bigl[ {\p\ln\rho\over \p x^\mu}
-\hbox{str}\, \wh\Gamma_\mu
\,-\,\hbox{str} \, \Bigl( {\p_l U^i_\mu\over \p \psi^j}
\Bigr) \, \Bigr]\, = 0\, , \eqn\yyeight$$
Using the covariant constancy  of the symplectic form
$D_\mu (\, \wh\Gamma\, ) \bra{\omega} = 0$ [\senzwiebach],
the condition that $F^i$ be a symplectic map reduces to the
condition that $U^i_\mu$ be a symplectic diffeomorphism.
This, in turn, implies that there is
an odd hamiltonian function $\bU_\mu$ such that
$$U^i_\mu  = \{\, \psi^i\, ,\, \bU_\mu\,\}\, \leftrightarrow \,
U^i_\mu = \omega^{ij} {\p_l \bU_\mu\over \p\psi^j} \, .\eqn\symdiff$$
This implies that \yyeight\ can be written as
$$\hbox{D}_\mu(\,\wh\Gamma\,) \, S\,-\,
\bigl( \hbar\Delta  \bU_\mu\, + \{ \,S \,, {\bf U}_\mu \,\} \bigr)\,=\,
\half\, \hbar\, \Bigl[\, \hbox{str}\, \wh\Gamma_\mu
\, - {\p\ln\rho\over \p x^\mu}\,\Bigr]\, .
\eqn\xyeight$$
This is our final form for the condition of local quantum background
independence. Notice that the right hand side contains only field
independent terms, and only at first power of $\hbar$.
Proving quantum background independence will amount to
finding an
odd hamiltonian  $\bU_\mu$ such that Eqn.\xyeight\ holds.
The diffeomorphism implementing background independence will then
read
$$\ket{\Psi}_{x+\delta x} = {}_{x+\delta x}{\cal I}_x \,\Bigl[\,
\ket{\Psi} - \delta x^\mu \,(\, \wh\Gamma_\mu \ket{\Psi} +
\ket{U_\mu}\, ) \,\Bigr]\, ,\eqn\dibi$$
where $\ket{U_\mu}\equiv \ket{\Phi_i}\, U^i_\mu$.
Finding the odd hamiltonian $\bU_\mu$ is our aim.

\section{The Conditions of Background Independence in
Geometric Language}

In this subsection we shall first rewrite the condition \xyeight\ of background
independence in a geometric language, and then separate out
the terms carrying different powers of $\hbar$ and
the string field $\ket{\Psi}$.
As we shall see in the later sections,
while the study of condition \xyeight\ will be efficiently carried out
in geometrical terms, the field independent terms at $\O(\hbar)$ will
require special attention.

We start with
the observation that the part of ${\bf U_\mu}$ which is
$\hbar$ independent and is linear in $\ket{\Psi}$,
should give rise to a constant
shift in $\ket{\Psi}$ proportional to $\ket{\wh\O_\mu}$, and, as in
Ref.[\senzwiebach], is given by ${\bf U}_{\mu(0,2)}$ defined in \shftpsi.
Using the insight obtained from the analysis of Ref.[\senzwiebach], we
start with the following ansatz for ${\bf U_\mu}$:
$${\bf U}_\mu = {\bf U}_{\mu\,(0,2)} - f_\mu (\B)\,,\eqn\ansatzb$$
where,
$$\B \equiv  \sum_{ g,n}\hbar^g \B_{g,n}  \quad\hbox{with}\quad
\cases{\B_{0,n}\equiv 0\,\,\hbox{for}\,\,n\leq 2,\cr
\B_{g,n}\equiv 0 \,\,\hbox{for}\,\,
n\leq 1\,.}\,\eqn\dfnvsp$$
Each space $\B_{g,n}$ will be a $(6g+2n-6)+1$ dimensional subspace of
$\wh\P_{g,n}$ (in our notation we should write $\B_{g,n}^{(1)}$, but we
will omit the superscript, as $\B$ spaces will always have one real dimension
more than the corresponding moduli space). The function $f_\mu$ was
defined in Eqn.\fgeomty.  $\B$ spaces will always have
one special puncture (for inserting $\ket{\wh\O_\mu}$) and will be symmetric
in all other punctures. Since any
field independent term in the hamiltonian ${\bf U}_\mu$ would drop out of
the background independence condition, the $\B$ spaces will
have at least one puncture besides the special one, where the string
field will be inserted. This is reflected in the range of $n$ given in
Eqn.\dfnvsp.
The proof of background independence now amounts to showing the
existence of the subspaces $\B_{g,n}$ such that Eqn.\xyeight\ is
satisfied. Note that $\B_{g,2}$ for $g\ge 1$ determine the quantum
corrections to the classical shift generated by ${\bf U}_{\mu(0,2)}$.

We now evaluate various terms in Eqn.\xyeight. We have
$$ \Delta{\bf U}_\mu= \Delta {\bf U}_{\mu(0,2)} - \Delta f_\mu (\B)\,
= f_\mu (\Delta \B)\,, \eqn\computfp$$
where use was made of Eqn.\hrehn. Furthermore, since $S_{1,0}$ is field
independent,
$$\eqalign {
\{ S , {\bf U}_\mu \} &= \,\{\, S_{0,2} + f(\V) \,,\, {\bf U}_{\mu (0,2)} -
f_\mu (\B) \, \}\,, \cr
&=-\{ S_{0,2}\, , f_\mu (\B) \} + \{ f(\V) \,,  {\bf U}_{\mu (0,2)}\}
- \{ f (\V) \,,\, f_\mu (\B) \}\,, \cr
&=\,f_\mu \bigl( \p\B + \underline{\V} + \{ \V , \B \} \bigr)\,,
\cr}\eqn\computsp$$
where we made use of Eqns.\hrehn--\vnshcon.\foot{Recall that the
underline is there to remind us that the space has one special puncture.}
The two equations above,
together with Eqn.\fcovder\ can now be substituted into the background
independence condition \xyeight. We then obtain

$$f_\mu \Bigl( \V'_{0,3} + \K\V -\p\B - \hbar\Delta\B - \{\V,\B\}-
\underline{\V} \Big)   =\,
\half\, \hbar\, \Bigl[\, \hbox{str}\, \wh\Gamma_\mu
\, - {\p\ln\rho\over \p x^\mu}\, - 2{\p S_{1,0}\over \p x^\mu} \,
\Bigr]\, . \eqn\mstsat$$

\noindent
This is the final geometrical version of the conditions for
background independence.
\subsection{Power Series Expansion}
We can now demand that the above equation holds separately for terms
containing different powers of $\hbar$ and $\ket{\Psi}$.
The field independent, order $\hbar$ contribution to the above equation
takes the form:
$$\p_\mu S_{1,0} =
\half\, \Bigl[\, \hbox{str} (\,\wh\Gamma_\mu\, ) - {\p \ln \rho\over \p
x^\mu}\, \Bigr]
+ f_\mu(\Delta\B_{0,3}) +
f_\mu(\underline{\V}_{1,1})
\, .\eqn\ebivac$$
Field independent terms of order $\hbar^g$ ($g\ge 2$) give
$$ f_\mu \Big( \K\V_{g,0} - \Delta\B_{g-1, 3} - \sum_{g_1=1}^{g-1}
\{\V_{g_1, 1},\B_{g-g_1,2}\}- \underline{\V}_{g,1} \Big) =0 \, .
\eqn\efieldindependentohtpg $$
Finally, order $\hbar^g \ket{\Psi}^N$ terms give,
$$ f_\mu(\V'_{g,N+1} -\p\B_{g, N+1} -\underline{\V}_{g, N+1}) =0\, ,
\quad \cases{ N\ge 2,\, \hbox{for} \, g=0\cr N\ge 1, \, \hbox{for} \, g\ge 1,}
\eqn\efielddependent $$
where,
$$ \V'_{g,N+1}= \K\V_{g,N} -\sum_{g_1=0}^g \sum_{m=1}^{N+1}
\{\V_{g_1,m}\, , \, \B_{g-g_1,N-m+3}\}
-\Delta\B_{g-1,N+3} \, \,
\cases{N\ge 3, \, \for \, g=0 \cr N\ge 1,\, \for\, g\geq 1\, .}
\eqn\epppsix$$
Note that we have specified to the right of the equation the restrictions
on values of $N$ for this equation to hold. At genus zero, $N\geq 3$ because
$\V'_{0,3}$ was already defined, and Eqn.\epppsix, for $N=2$, would give
$\V'_{0,3}=0$.

Looking at these equations, it is clear that the conditions
\efieldindependentohtpg\ and \efielddependent\ are purely geometrical
in the sense of Riemann surfaces,
whereas the condition \ebivac\ involves extra theory space elements.
In the next
section we shall show how to find the spaces $\B_{g,n}$ satisfying
Eqns.\efieldindependentohtpg\ and \efielddependent, leaving the analysis
of Eqn.\ebivac\  to \S6.

\chapter{Background Independence of $d\mu_S$ up to $\hbar$-Independent
Normalization}

In this section we will establish the background independence of
$d\mu_S$ up to an overall $\hbar$ independent normalization factor.
This will involve finding subspaces $\B_{g,N+1}$ of $\wh\P_{g,N+1}$
satisfying Eqns.\efieldindependentohtpg\ and \efielddependent. In fact, we
shall first solve Eqns.\efielddependent\ to find $\B_{g,N+1}$, and then
show that these $\B_{g,N+1}$ automatically satisfy
Eqn.\efieldindependentohtpg.

We shall now carry out a
consistency check of the set of Eqns.\mstsat. Ignoring the order $\hbar$
constant terms, and using Eqn.\basicint, we write \mstsat\ as
$$\p\B \simeq \V' -\underline{\V}\,,\quad\hbox{with}\quad \V' \equiv
 \V'_{0,3} + \K\V - \hbar\Delta\B - \{\V,\B\}\,. \eqn\rwrt$$
Here the
symbol $\simeq$ is used to denote that the equality holds up to the
choice of local coordinate at the special puncture, {\it and} up to the
addition of terms of order $\hbar$ representing  surfaces of genus one
carrying only the special puncture,
but no ordinary puncture. Such surfaces will produce order $\hbar$ constant
terms when acted on by $f_\mu$.
If we now apply the $\p$ operator again on this equation, then the left
hand side of this equation vanishes. Thus for consistency, the right hand
side must also vanish up to order $\hbar$ terms which do not carry any
ordinary puncture.
The verification of this condition is straightforward. We have,
$$\p^2\B \simeq \p\K\V +\hbar\Delta\p\B-\{\p\V,\B\} + \{
\V,\p\B\}-\p\underline{\V}\, \eqn\sptro$$
Using Eqn.\rwrt\ for $\p\B$, and the various
identities derived in \S2, we find,
\foot{Note that we can use Eqn.\rwrt\ because the terms in $\p\B$ of order
$\hbar$ and only one (special) puncture, which we do not keep track, do not
contribute to the right hand side of Eqn.\sptro.}
$$\eqalign{\p^2\B &\simeq \,-\{ \K\V,\V\} - \hbar\Delta\K\V -
\{ \V'_{0,3},\V\} \cr
&\quad +\hbar (\Delta \V'_{0,3} + \Delta\K\V - \{\Delta\V,\B\}
+\{ \V,\Delta\B\} -
\Delta\underline{\V}) \cr
&\quad +\half \{ \{ \V,\V\},\B\} + \hbar\{ \Delta \V ,\B\} \cr
&\quad+ \{ \V\, ,\, \V'_{0,3} + \K\V - \hbar\Delta\B -
\{\V,\B\}-\underline{\V}\}\cr
&\quad+ \{ \underline{\V} ,\V \} + \hbar
\Delta\underline{\V}\, ,\cr} \eqn\themess$$
where, for the help of the reader, each term in \sptro\ has been written as
one line in \themess. Using the symmetry properties of $\{,\}$, and
the Jacobi identity,
$$\half \{ \{\V ,\V\}\, , \, \B\} - \{\V\, , \, \{\V,\B\}\} =0 \, ,
\eqn\enewjacobi $$
it is easily checked that all terms in the right
hand side cancel out. The term $\hbar\Delta \V'_{0,3}$ is of order $\hbar$ and
does not carry any ordinary puncture; hence it
must be dropped in the
present calculation.
This finishes the verification of the consistency condition. The analysis
in the later part of this section will involve showing how, using this
consistency condition, we can construct the vertices $\B_{g, N+1}$
following a recursive procedure.

\section{Background Independence of the Field Dependent Part of $d\mu_S$}

We now turn our attention to the construction of explicit solutions of
Eqn.\efielddependent. This equation
is satisfied, provided we find a set of vertices
$\B_{g,N+1}$ satisfying,
$$\p\B_{g,N+1}\approx \V'_{g,N+1} -\V_{g,N+1}\,, \eqn\eqqqfive$$
where $\approx$ denotes equality up to a choice of the coordinate system
at the special puncture.
In arriving at this equation we made use of \basicint.
The existence of $\B_{g, N+1}$ satisfying Eqns.\eqqqfive\ and \epppsix\
can be proved by induction as follows.
Let $b_{g,N}\equiv (6g+2N-6)+1$ denote the dimension
of the space $\B_{g,N}$.
We assume that the spaces $\B_{g,N+1}$ satisfying
Eqns.\eqqqfive\ and \epppsix\ have been constructed
for all $g$ and $N$ satisfying $1\le b_{g,N+1}\le b_0$
for some odd $b_0\ge 1$,
and then show that $\V'_{g_0,M+1}$  defined through
Eqn.\epppsix\ satisfies the consistency condition
$$\p\V_{g_0,M+1}'\approx\p\V_{g_0,M+1}\,,  \eqn\equantcons$$
for
$b_{g_0, M+1}=b_0+2$. (Remember that $b_{g,N}$ is always odd.)
Once Eqn.\equantcons\ is satisfied, the space of surfaces
$\B_{g_0,M+1}$
can be defined as a symmetric homotopy between
$\V_{g_0,M+1}$ and $\V_{g_0,M+1}'$. \foot{See Ref.[\senzwiebach] for a
canonical procedure to construct a symmetric homotopy.}
This allows
us to define $\B_{g_0 , M+1}$ satisfying Eqn.\eqqqfive\ for all $g_0$ and $M$
for which $b_{g_0,M+1}=b_0+2$.
Equation.\equantcons\ is satisfied trivially for $b_{g, N+1}=1$.
Indeed, this case corresponds to $g\hskip-2pt=0,M\hskip-2pt=2$, and both
$\V'_{0,3}$ and $\V_{0,3}$ have zero dimension, and therefore zero boundary.
The lowest dimensional $\B$ space is therefore $\B_{0,3}$ which is a symmetric
homotopy between $\V_{0,3}$ and $\V'_{0,3}$.
The induction argument will then
imply that consistent $\B_{g,N}$'s
can be constructed for all values of $N\ge 3$ for $g=0$,
and all values of $N\ge 2$ for $g\ge 1$.

Thus the task that remains is to
calculate $\p(\V'-\underline{\V})_{g_0,M+1}$ using
the expression \epppsix\ for $\V'_{g_0, M+1}$, and show that it vanishes.
This expression contains
$\p\B_{g, n}$ for different $g$ and $n$ on the right hand side, but all
of these  $\B_{g,n}$'s
are of dimension $\le b_{g_0,M+1}-2=b_0$. Hence we can replace these
$\p\B_{g,n}$'s by $\V'_{g,n}-\V_{g,n}$, with $\V'_{g,n}$ defined
through Eqn.\epppsix.
It is clear, however that the resulting expression is precisely equal to
the term on the right hand side of Eqn.\themess\ which is
of order $\hbar^{g_0}$
and carries a total of $M+1$ punctures (including the special puncture).
We have already shown that the right hand side of Eqn.\themess\ vanishes
except for a $g_0=1$, $M=0$ term. This, in turn, shows that our expression
indeed vanishes as long as $(g_0,M)\ne (1,0)$.
This condition is automatically satisfied, since $\B_{g_0, M+1}$ is
defined only in the range
$M\ge 2$ for $g_0=0$, and $M\ge 1$ for $g_0\ge 1$.
This completes the proof by induction, and the complete construction
of the $\B$ spaces.

\section{Background Independence of the $\hbar$-Dependent
Normalization of $d\mu_S$}

In this subsection we verify that the $\B_{g,N+1}$, constructed in the
previous subsection, does satisfy Eqn.\efieldindependentohtpg.
This equation can be rewritten as,
$$f_\mu(\V'_{g,1} -\underline{\V}_{g,1}) =0\,, \eqn\rewriting $$
where, as the notation suggests,
$$ \V'_{g,1} \equiv \K\V_{g,0}- \sum_{g_1=1}^{g-1}\{ \V_{g_1,1}\, , \,
\B_{g-g_1, 2} \}
- \Delta \B_{g-1, 3}\,\,, \eqn\evacuumsix $$
is the subspace of $\V'$ of genus $g$ and one special puncture.
In order to satisfy Eqn.\rewriting, it suffices if
$\V'_{g,1}\approx\V_{g,1}$.
By definition, this is equivalent to $\pi_f\,(\V_{g,1}) = \pi_f\,(\V'_{g,1})$,
where $\pi_f$ is now the projection  from $\wh\P_{g,1}$ to $\M_{g,1}$.
Both $\pi_f(\V_{g,1})$ and $\pi_f(\V'_{g,1})$
are subspaces of $\M_{g,1}$ of codimension 0.
Furthermore, neither contains degenerate surfaces. This implies
that $\pi_f\,(\V_{g,1}) = \pi_f\,(\V'_{g,1})$ will hold
provided
that $\p\,\pi_f(\V_{g,1}) = \p\,\pi_f(\V'_{g,1})$.
This last condition is equivalent to
$\pi_f\, (\p\V_{g,1}) = \pi_f\,(\p\V'_{g,1})$, which
by definition is equivalent to $\p\V_{g,1} \approx  \p\V'_{g,1}$.
Thus, if we can show that,
$$\p\,(\V'_{g,1} -\V_{g,1})\approx0\,, \eqn\uhuh$$
we would have established background independence of the order $\hbar^{g-1}$
constant terms in the action-weighted measure.
Using the fact that all $\B$ spaces satisfy Eqns.\eqqqfive\ and \epppsix,
it is easy
to verify that the left hand side of the above equation can be identified
to the order $\hbar^g$ terms on the right hand side of Eqn.\themess\
which do not carry any ordinary puncture. We have already seen that this
vanishes for $g\ge 2$. This establishes background independence of
order $\hbar^{g-1}$ field independent terms in the action weighted measure
for all $g\ge 2$.

\chapter{Background Independence of the $\hbar$-Independent
Normalization of $d\mu_S$}

We shall now turn to the analysis of the terms that
control the $\hbar$-independent part of the normalization of the
physical measure $d\mu_S$. There are two ingredients to this normalization.
One is the genus one part $S_{1,0}$ of the master action. As pointed out
earlier, there is no {\it a priori} geometric construction of $S_{1,0}$,
and the problem here
is to construct an appropriate $S_{1,0}$ using the requirement of
background independence. The other ingredient is the scale factor
$\rho(x)$ of the measure (Eqn.\selement). Since both $S_{1,0}$ and $\rho(x)$
 determine the $\hbar$ independent part of the
normalization of the measure $d\mu_S$, we need to discuss the
construction of $S_{1,0}$ and $\rho(x)$ together.
We begin  with some preliminary observations relevant to
the problem at hand. Then, we discuss the relevant equations, and
determine both $S_{1,0}$ and $\rho(x)$. Our determination of $\rho(x)$
is somewhat indirect, and is based on the background independence of
the free energy of string field theory.

\section{Preliminary Observations}

\subsection{Basis Independence.}
We may ask if observables are
affected by a change of basis in $\H$.
It is clear that observables
will be basis independent if the construction of
closed string field theory is fully basis independent.
The string field can be written as
$$\ket{\Psi} = \sum \ket{\Phi_i} \, \psi^i \,=\, \sum \ket{\Phi'_i} \, \psi'^i
\, ,\eqn\twobasis$$
where we have used two different basis, related by some invertible
transformation  $\ket{\Phi'_i}= \ket{\Phi_j}A^j_{~i}$.
Basis independence of the string field theory requires that
$d\mu_S (\psi)$ and $d\mu_S (\psi')$ agree when $\psi^{\prime i} =
(A^{-1})^i_{~j}\psi^j$.
It is important to note that the construction of the string field action
is completely basis independent, since the string field $\ket{\Psi}$ is
basis independent and the string field vertices $\bra{V^{(g,N)}}$ are also
basis independent. Our choice of measure $d\mu$ given by
$d\mu=\rho(x)\prod d\psi^i$ with some fixed $\rho(x)$, appears to be
basis dependent. In order to keep $d\mu$ basis independent, $\rho(x)$
must transform under a change of basis: $\rho(x)\to \rho(x)\, {\rm
sdet}(A)$. Operationally, we choose a basis in $\H$, determine a
suitable scale factor $\rho(x)$, and formulate string field theory.
If another
basis is desired, one must readjust $\rho(x)$ using the condition of
invariance of the measure.
Interpreted this way, the physical action weighted
measure $d\mu_S\equiv(\rho(x)\prod d\psi^i_x)\exp (2S_x/\hbar)$ is
manifestly basis independent.

\subsection{What can be Determined?}
In separating the measure factor $\rho$ from $S_{1,0}$ there is an ambiguity.
The term $S_{1,0}$ of the string field action enters in $d\mu_S$
as an overall ($\hbar$ independent) constant. Thus,
we could, in principle, absorb $\rho(x)$ into $S_{1,0}$ in the form
$S_{1,0}=\half\ln\rho(x)+ \cdots$ where the dots would indicate basis
independent terms. We will not do so. It is natural
to demand that $S_{1,0}$
be a scalar, since it is part of the action, which in the BV formulation
is taken to be a scalar. This requirement, of course, is not enough
for splitting unambiguously the scale factor between $\rho$ and
$S_{1,0}$, as we can always shift scalar factors back and forth
between them. In satisfying the demands of background independence,
we will be able to take for $S_{1,0}$ a Riemann surface geometrical
object, very much as we did for the higher vacuum vertices $S_{g,0}$
($g\ge 2$). We will then get an equation for $\p_\mu\ln\rho$
involving a theory space connection. This equation encodes naturally
the required basis dependence of $\rho$.
Working in any given basis, the overall $x$-independent multiplicative
factor of $\rho$ is not determined by the condition of background
independence. We must use an observable to fix this constant.
The one-loop free energy of string field theory
is the appropriate observable.
It is guaranteed by the master equation to be gauge independent.
Working at some specific point $x_0$ in CFT theory space, we demand that
$\rho(x_0)$ be chosen in such a way that the one loop free energy
from string field theory  around $x_0$ is equal to the partition
function of CFT$_{x_0}$, integrated over one copy of the full moduli space
of the torus.\foot{It is best to do it in a basis in which the one
loop vacuum graph is finite up to tachyon divergence,
since in such a basis $\rho(x_0)$ will also
be finite. Later in this section we shall discuss how this can be done.}
The same result would then hold for string field theory formulated
around any CFT in the space as a consequence of background
independence of string field theory, and the
result[\ghoshalsen] that the difference in one loop free energy of string field
theory expanded around two nearby classical solutions correctly accounts for
the change in CFT partition function under a  marginal deformation.

\subsection{On Non-canonical and Canonical Transformations}
String field theory can be formulated with
stubs on all the vertices and with a standard kinetic term, or alternatively
it can be written with a cutoff propagator, by adding factors of
$\exp (2aL_0^+)$
in the kinetic term, together with vertices that have no stubs. If
both formulations use the same measure, they
yield exactly the same observables for processes involving external legs.
Nevertheless they do not yield the same vacuum graph; using the same
basis and the same gauge, it is clear that the
respective propagators have different traces.
This appears to be a somewhat surprising result at the first sight
because the two actions can easily be seen to be
related by a field redefinition of the form $\psi^i\to exp(a\gamma_i) \psi^i$,
where $\gamma_i$ is the conformal dimension of the state $\ket{\Phi_i}$.
Since the fields and antifields, which have opposite statistics, are
scaled by the same amount, one gets ${\rm sdet}(A)=1$ under this
transformation, and hence no change in $\rho$. In order to understand the
origin of the change in the vacuum graph, we note that the transformation
$\psi^i\to exp(a\gamma_i)\psi^i$ is not canonical, since the fields and the
antifields are not scaled in appropriately correlated way. Hence
if we want this transformation to generate an equivalent theory, we must
transform the symplectic form $\omega$ in addition to the action.
This change in $\omega$ gives rise to an extra factor in the construction
of $d\lambda$ from $d\mu$ using Eqn.\LAMBDA, and compensates for the
change in the vacuum graph.
In contrast, we
note that a change of cell decomposition in closed string field theory,
induced by changing the three string and higher string vertices, is always
implemented by a {\it canonical} transformation [\hatazwiebach]. This
transformation left all observables, including
one loop free energy, unchanged.
Under an infinitesimal canonical transformation generated by the
hamiltonian $\epsilon$, we have that $d\mu \exp(2S/\hbar)\to
d\mu (1+ 2\Delta \epsilon ) \exp\bigl( 2[S+ \{ S,\epsilon \}]/\hbar\bigr)$.
In Ref.[\hatazwiebach] the genus zero contribution to $\epsilon$ was
cubic in the string field and therefore
$\Delta\epsilon$ cannot give an $\hbar$ independent and field
independent contribution. Nor can $\{ S,\epsilon\}$, since the
classical action is at least quadratic in the string field.  The
invariance of the $\hbar$-independent normalization of the measure is
consistent with the fact that the kinetic term is unchanged under the
canonical transformation, which, in turn, leaves the one loop vacuum graph
unchanged. The one loop free energy, which is a sum of the vacuum graph
and half of the logarithm of the $\hbar$ independent normalization of
the measure, is then left unchanged, as it should be.

\section{Equations for $S_{1,0}$ and $\ln\rho$}

Now that we have given an overview of most of the relevant issues we
can proceed to the analysis of equation \ebivac, which reads
$$\p_\mu S_{1,0}   = -\half\,{\p\ln\rho\over \p x^\mu}\,+\, \half\,
\hbox{str}\,(\wh\Gamma_\mu) + f_\mu(\Delta\B_{0,3})
+f_\mu(\V_{1,1}) \,,\eqn\exonex $$
As written, Eqn.\exonex\ is not manifestly free of divergences. For
example, the term $f_\mu(\Delta\B_{0,3})$
involves integration over the space
$\Delta\B_{0,3}$. Since
$\B_{0,3}$ (recall $\p\B_{0,3} = \V'_{0,3}-\V_{0,3}$)
is an interpolation from the closed string vertex
$\V_{0,3}$ to the auxiliary vertex $\V'_{0,3}$, and
$\Delta\V'_{0,3}$ gives singular tori,
the integral above is potentially
divergent. Also $\hbox{str}\, \wh\Gamma_\mu$ is potentially divergent.
We will now show  that with an appropriate choice of basis,
$\half \hbox{str}\,(\wh\Gamma_\mu) + f_\mu(\Delta\B_{0,3})$ can be made finite.
To this end we introduce a
connection $\wt\Gamma$, defined through the relation:
$$ \bra{\omega_{12}} (\wt\Gamma_\mu)^{(1)} =  \bra{\omega_{12}}
(\wh\Gamma_\mu)^{(1)}
+\int_{\wt\B_{0,3}} \bra{\Omega^{(1)0,3}}\wh\O_\mu\rangle\, ,
\eqn\edefwtgamma$$
where $\wt\B_{0,3}$ is a path in $\wh\P_{0,3}$ interpolating between
$\V'_{0,3}$ and another point $\wt\V_{0,3}$:
$$\p\wt\B_{0,3} =\wt\V_{0,3} - \V'_{0,3} . \eqn\edefbthree $$
The new auxiliary vertex $\wt\V_{0,3}$  is taken to be a sphere
punctured at $z=0$ with local coordinate $w_1=e^a z$, at $z=\infty$, with
local coordinate $w_2=e^a/z$, and at $z=1$, with a local coordinate $w_3$
which we shall keep arbitrary for the time being.
The constant $a$ is some number $\ge 0$.
We can now use Eqns.\fgeomty\ and \edefwtgamma\ to write
$$\eqalign{
\half\,\hbox{str}\,(\wh\Gamma_\mu) + f_\mu(\Delta\B_{0,3})
=  &
\half\, \hbox{str}\,(\,\wt\Gamma_\mu )\, - \half \hskip-4pt\int_{\wt\B_{0,3}}
\bra{\Omega^{(1)0,3}} \s\rangle \cdot \ket{\wh\O_\mu} +
f_\mu(\Delta\B_{0,3})\cr
= & \half\,\hbox{str}\,(\,\wt\Gamma_\mu )\, +
\hskip-6pt\int_{\Delta(\B_{0,3}+\wt\B_{0,3})}
\bra{\Omega^{(0)1,1}} \wh\O_\mu\rangle \, .\cr} \eqn\trcon$$
where we have used ${\rm str}\, A=\bra{\omega_{12}}A^{(1)}\ket{\s_{12}}$,
which holds for any operator $A$.
(The superscript ${}^{(1)}$ denotes that the operator acts on the first
state space.)
Here  $\B_{0,3}+\wt\B_{0,3}$ is a path from
$\V_{0,3}$ to $\wt\V_{0,3}$ made by joining together the path
$\B_{0,3}$ going from $\V_{0,3}$ to $\V'_{0,3}$, and the path
$\wt\B_{0,3}$ going from $\V'_{0,3}$ to $\wt\V_{0,3}$. The crucial
point is that the integral above is independent of the path we take! For two
paths $\C_{0,3}$ and $\C'_{0,3}$ with identical endpoints, so that
there is a $\D_{0,3}$
with $\p\D_{0,3} = \C_{0,3}-\C'_{0,3}$, we have from Eqn.\eketbrst\
$$\eqalign{
\Bigl(\, \int_{\C_{0,3}}-\int_{\C'_{0,3}}\,\Bigr)
\bra{\Omega^{(1)0,3}}
\s_{12}\rangle\cdot\ket{\wh\O_\mu}_3 &= \int_{\p\D_{0,3}}
\bra{\Omega^{(1)0,3}} \s_{12}\rangle \cdot
\ket{\wh\O_\mu}_3 \cr &= \int_{\D_{0,3}}\bra{\Omega^{(2)0,3}}
(Q^{(1)}+Q^{(2)}+Q^{(3)})\ket{\s_{12}}\cdot\ket{\wh\O_\mu}_3= 0\,,\cr}
\eqn\nochange$$
since $(Q^{(1)}+Q^{(2)})\ket{\s_{12}}=0$, and $Q\ket{\wh\O_\mu}=0$.
Path independence allows us to replace in \trcon\ the problematic
path $(\B_{0,3}+\wt\B_{0,3})$ by a new path $\B'_{0,3}$ from
$\V_{0,3}$ to $\wt\V_{0,3}$ which avoids $\V'_{0,3}$
$$\p \B'_{0,3} = \wt\V_{0,3} - \V_{0,3}\,.\eqn\newpathc$$
We choose this path
such that for every surface in this path, the $\Delta$ operator gives
a regular torus.
This is guaranteed if the coordinates curves of the symmetric punctures do
not touch. Since this is the case both for $\V_{0,3}$ and $\wt\V_{0,3}$
there is clearly a homotopy satisfying this condition.
This shows that the second term on the right hand side of Eqn.\trcon\
is finite.\foot{Since our proof of path independence assumes
finiteness of the quantities involved, we should interpret our
construction as a way to make manifest a cancellation of infinite
quantities (regularization). There should be a way to eliminate every
reference to $\V'_{0,3}$ from the start.}
While this term is basis independent, the first term  ${\rm
str}(\wt\Gamma_\mu)$ is basis dependent, and can be made finite by
an appropriate choice of basis. In particular, if the basis states are
chosen such that they are parallel transported by the connection
$\wt\Gamma_\mu$, then in this basis $\wt\Gamma_\mu$ vanishes, and hence
${\rm str}(\wt\Gamma_\mu)=0$.\foot{It is always possible to arrange that the
basis states at $x+\delta x$ are related to the basis states at $x$ via
parallel transport by the connection $\wt\Gamma_\mu$ to first order in
$\delta x$. This is all we need to set $\wt\Gamma_\mu=0$ at $x$.}
We shall not choose basis states at this stage.
Equation \exonex, with the help of \trcon\  can now be written as
$$  \p_\mu S_{1,0}\, - \hskip-6pt\int_{\V_{1,1}+ \Delta\B'_{0,3}}
\hskip-6pt\bra{\Omega^{(0)1,1}}\wh\O_\mu\rangle = \half\,
\Big(\hbox{str}\,(\,\wt\Gamma_\mu\,) \,-\,
\p_\mu \ln\rho\,\Big)\,\, .
\eqn\eoneloopthree $$
We will solve this equation by requiring both
$$ \p_\mu S_{1,0} = \int_{\V_{1,1}+ \Delta\B'_{0,3}}
\bra{\Omega^{(0)1,1}}\wh\O_\mu\rangle \, , \eqn\ejjjone$$
and,
$$ \p_\mu\ln\rho = \hbox{str} (\wt\Gamma_\mu) \, . \eqn\ejjjtwo$$
We must now show that the above two equations can be solved for
$S_{1,0}$ and $\ln\rho$.

\section{Explicit Determination of $S_{1,0}$}

Let us first discuss Eqn.\ejjjone.
Since the position of the puncture in a torus is irrelevant, the set of
surfaces $\V_{1,1}+ \Delta\B'_{0,3}$, determines a
set of surfaces $\V_{1,0}\subset \M_{1,0}$ by the simple operation of
deleting the puncture
$$\V_{1,0} \equiv \pi_F \,(\V_{1,1}+ \Delta\B'_{0,3})\, ,\eqn\gzerovac$$
where
$\pi_F$ denotes the operation of forgetting the puncture.
We now claim that
$$S_{1,0} = \int_{\V_{1,0}}\Omega^{(0)1,0} \, ,\quad\hbox{with}\quad
\Omega^{(0)1,0}_\Sigma\equiv {d\tau_1\wedge d\tau_2\over
\tau_2}\, Z^G Z^M_x(\tau, \bar\tau) \, ,\eqn\stansts$$
where we have defined the volume two-form $\Omega^{(0)1,0}_\Sigma$
(on $T_\Sigma(\M_{1,0})$) where the surface $\Sigma$ is a torus with
modular parameter $\tau$.
Here $Z^G\equiv Tr_{\rm ghost}\big( (-1)^{n_G} e^{2\pi i \tau L_0^G}
e^{-2\pi i \bar\tau \bar L_0^G} b_0 c_0 \bar b_0 \bar c_0\big) =
(\eta(\tau))^2 (\ov{\eta(\tau)})^2$ is the ghost
partition function (computed after
removing the ghost zero modes on the torus), and
$Z^M_x(\tau, \bar\tau)$ denotes the partition function of the matter CFT
at some point $x$. It follows from the analysis of ref.[\ghoshalsen] that
$$\p_\mu Z^M_x =\,-\,{\tau_2\over\pi} \vev{\O_\mu}^M_{g=1}, \quad
\p_\mu Z^G=0, \eqn\epartitwo$$
where $\vev{\O_\mu}^M_{g=1}$ denotes the one point function of $\O_\mu$ on
the torus in the matter CFT. In computing this correlation function, the
local coordinate at the point of insertion of the marginal operator
$\O_\mu$ on the torus must be taken to be the natural complex coordinate
$w$ on the torus, in terms of which the torus looks like a parallelogram
with the identification $w\equiv w+1\equiv w+\tau$. We can now compute
$\p_\mu S_{1,0}$ using \stansts\ and \epartitwo. We immediately find
$$ \p_\mu S_{1,0} =\,-\,{1\over\pi} \int_{\V_{1,0}}
d\tau_1\wedge d\tau_2\,\, Z^G
\vev{\O_\mu}^M_{g=1}\,\,. \eqn\epasss$$
It will be shown in appendix B that
$${1\over\pi} \int_{\V_{1,0}} d\tau_1\wedge d\tau_2\,\, Z^G
\vev{\O_\mu}^M_{g=1}\,\, = \,-\int_{\V_{1,1}+\Delta\B_{0,3}'}
\bra{\Omega^{(0)1,1}}\wh\O_\mu\rangle\, . \eqn\epartithree$$
The last two equations imply that our ansatz for $S_{1,0}$ (Eqn.\stansts)
satisfies the condition given in Eqn.\ejjjone.
We point out that, as desired, $S_{1,0}$ is a geometrical basis independent
object (a scalar on $\H$).
We can get to understand what is the region of integration $\V_{1,0}$ of the
genus one vacuum vertex defining $S_{1,0}$ by computing the boundary of
$\V_{1,0}$. It follows from \gzerovac\ that
$$ \p\V_{1,0} = \pi_F \,(\p\V_{1,1}- \Delta\p\B'_{0,3})\, =
 -\, \pi_F \, \Delta\wt\V_{0,3} \,,\eqn\eoneloopeight $$
where use was made of Eqns.\propvgn\ and \newpathc.
Since $\wt\V_{0,3}$ has stubs of length $a$ in the coordinate system in
which the string has length $2\pi$,
this means that $S_{1,0}$  is given by the integral
of the partition function over $\M_{1,0}-R$, where $R$,
in the usual representation of $\M_{1,0}$ in the $\tau$ plane, is
the region $-\half \leq \tau_1 <\half\,,\,\,$
$\tau_2 >{a\over \pi}$.

\section{Determination of $\rho$}

We must now discuss the equation $\p_\mu \rho = \hbox{str} (\wt\Gamma_\mu )$.
This equation is sensible as both sides of the equality
transform in the same way under a change of basis.
It is also intuitively clear why the
theory dependence of $\rho$ should be affected by a connection implementing
background independence. After expanding an SFT around a nontrivial background
we get a theory that gives the same observables as the SFT formulated
on a CFT corresponding to
that background. Nevertheless to relate the expanded theory to the
theory on the second background we need to perform, to first order, a
change of basis encoded by a connection. Thus properties of this connection
must arise in comparing the measures.

It would be desirable to give a direct proof that
$\p_\mu \rho = \hbox{str} (\wt\Gamma_\mu )$ is an integrable equation.
This would involve showing that $\hbox{str}
(\p_\mu \wt\Gamma_\nu - \p_\nu \wt\Gamma_\mu) =0$. We are not quite able
to do this at present.
An argument that the equation is integrable can be given if we
are cavalier about divergences. From Eqn.\edefwtgamma\ we have that
$$ \hbox{str}(\,\wt\Gamma_\mu) =\hbox{str}(\wh\Gamma_\mu)
-2\int_{\Delta\B_{0,3}}\bra{\Omega^{(0)1,1}}\wh\O_\mu\rangle\, .
\eqn\argure$$
Since the connection $\wh\Gamma$ is flat, we can choose a basis in
which it vanishes in a neighborhood of a point $x_0$.
(There are of course problems with infinities here).
By the result of appendix B, the second term is given by
$\bigl( - 2\p_\mu \int_{\pi_F(\Delta\B_{0,3})}
\Omega^{(0)1,0}\bigr)\,.$
Hence $\hbox{str}(\wt\Gamma_\mu)$ is derivative of a quantity, making the
integrability of the $\rho$ equation clear (this quantity, however,
is divergent due to the presence of singular tori in $\Delta\B_{0,3}$).
This argument suggests that if we could regulate properly everything
we should not have problems establishing integrability. What we will do next,
however, is to find $\rho(x)$ directly from the condition of background
independence as it applies to observables. The observable in question
will be the free energy of string field theory.

Let  SFT$_x$ denote the closed string field theory formulated around CFT$_x$.
Moreover, let $F_{x}(y)$ denote the free energy calculated from
SFT$_{x}$ expanded around a nontrivial classical solution representing
CFT$_y$. In this notation $F_{x} (x)$ is simply the free energy
calculated from SFT$_{x}$ using the trivial background $\ket{\Psi_{x}}=0$.
The computation of $F_x(x)$ involves computing a one
loop vacuum graph, and including appropriately the measure factor $\rho$
and the constant $S_{1,0}$. We work in the Siegel gauge, which is fixed as
follows.
We split the full set of string fields that appear in the BV
formalism into two groups.  Let
$\{\ket{\Phi^{(n)}_r}\}$ denote a set of basis states in $\H$
of ghost number $n$ and annihilated by
$b_0^+$, and $\{\ket{\wt\Phi^{(n)}_r}\}$ denote the set of basis states
in $\H$ of ghost number $n$ and
annihilated by $c_0^+$, normalized in such a way that
$$\bra{\omega_{12}\,}\Phi^{(m)}_s\rangle_{(1)}\ket{\wt \Phi^{(5-n)}_r}_{(2)}
 = \delta_{rs}\delta_{nm}\,  .\eqn\ebasisnormalization$$
(This choice of basis still leaves freedom to be exploited later.)
If we denote by $\psi^r_{(n)}$ and
$\wt\psi^{r}_{(n)}$ the coefficients of $\{\ket{\Phi^{(n)}_r}\}$ and
$\{\ket{\wt\Phi^{(n)}_r}\}$ in the expansion of the string field, then
with the choice of normalization we have given, the pair
$\bigl( \psi^r_{(n)}\, ,\wt\psi^{r}_{(5-n)}\bigr) $ is a
properly normalized field/antifield pair. The original BV
measure now can be expressed as $\rho(x)\prod_\alpha d\psi^\alpha
d\wt\psi^{\alpha}$ where the index $\alpha$ runs over both $r$ and $n$. The
Lagrangian submanifold is now taken to be the surface $\wt\psi^{\alpha}=0$.
{}From Eqn.\LAMBDA\ we get $d\lambda = \sqrt{\rho(x)} \prod_\alpha
d\psi^\alpha$. The free energy of SFT$_x$ is now given by,
$$\eqalign{ \exp \bigl( F_x(x)\bigr) \, =& \,\sqrt{\rho(x)}\,
\prod_\alpha d\psi^\alpha
\exp\,\Bigl( {1\over \hbar} S\,(\psi^\alpha, \wt\psi^{\alpha}=0) \Bigr)\, ,\cr
=& \exp \Bigl( \,\half \ln \hbox{sdet} K(x) + \half\ln \rho(x)
+S_{1,0}(x)+\O(\hbar) \,\Bigr)\, , \cr}
\eqn\efreeenergy$$
where $K$ denotes the gauge fixed kinetic term matrix in the chosen basis
for the string field. The first term on the last line of the above
equation is the so called vacuum graph.

As has already been pointed out before, Eqn.\ejjjtwo\
leaves an overall multiplicative constant in $\rho$ undetermined.
To fix this constant we must specify a value for $\rho$ at some point $x_0$ in
the CFT theory space.
We will choose $\rho(x_0)$ in such a way that the free energy
$F_{x_0} (x_0)$ of the string field theory SFT$_{x_0}$
is equal to the partition function of
CFT$_{x_0}$ integrated over $\M_{1,0}$
$$F_{x_0} (x_0) = \int_{\M_{1,0}} Z_{CFT_{x_0}}\, .\eqn\obsfebb$$
In view of the last two equations, we have that
$$\ln\rho(x_0)= \Big(2\int_{\M_{1,0}} Z_{CFT_{x_0}} -\ln\hbox{sdet}K(x_0) -
2S_{1,0}(x_0) \Big) \, .\eqn\obsfree$$
The quantity $\ln\hbox{sdet}K(x_0)$ can be computed
for a given choice of basis, and  $S_{1,0}(x_0)$ is given
by Eqn.\stansts. Thus
we can use Eqn.\obsfree\ to fix $\rho(x_0)$.
Later we shall see how to choose the basis
states so that  $\rho (x_0)$ (and also $\rho(x)$) turns out to be finite.

It was proven in Ref.[\ghoshalsen], that Eqn.\obsfebb\ guarantees
that the free energy calculated with SFT$_{x_0}$, expanded around a classical
solution representing CFT$_x\,$, gives the integral of the partition function
of CFT$_x$:
$$F_{x_0} (x)= \int_{\M_{1,0}} Z_{CFT_{x}}\, .\eqn\obsfreeb$$
This analysis rests on two basic results, one giving the variation
of the partition function when we deform a CFT using a marginal
operator, and the other being the geometrical recursion relation
$\p \V_{1,1} = -\Delta \V_{0,3}$.

Background independence implies that the physics of SFT$_x$ can
be equally obtained from SFT$_{x_0}$ expanded around a classical
solution representing CFT$_x$. Since free energy is an observable
in string field theory, we must have that $F_x(x) = F_{x_0} (x)$,
and therefore, in view of the above equation
$$F_{x} (x) = \int_{\M_{1,0}} Z_{CFT_{x}}\, .\eqn\obsfreebb$$
Thus background independence demands that
not only at $x_0$, but for any $x$ in a neighborhood of $x_0$,
$\rho(x)$ must be chosen
so that $F_x(x)$ calculated using Eqn.\efreeenergy\ equals the right
hand side of the above equation. Thus we have to choose
$$\rho(x) = \wh\rho(x)  \equiv \,
\exp \,\Bigl[\, 2\,  \Bigl( \,\int_{\M_{1,0}} Z_{CFT_x} - S_{1,0}(x) -
\half\ln\hbox{sdet}K(x)\Bigr)\, \Bigr]\,.\eqn\erhoexpression$$
For a given choice of basis states around $x_0$, $K(x)$ is fixed, and
hence the above equation determines $\rho(x)$ in a finite neighborhood
of $x_0$ uniquely. \foot{The one loop vacuum
graph and the integral of CFT partition function over $\M_{1,0}$, have tachyon
divergences. These divergences cancel out and do not affect the
determination of $\rho$.}

We now argue that the above choice of $\rho$ is also sufficient to ensure
background independence, in particular it must satisfy the required
equation $\p_\mu\ln\wh\rho=\hbox{str}(\wt\Gamma_\mu)$.
Let $x$ be any point in
the CFT theory space, and $n^\mu$ be an arbitrary tangent vector at $x$.
Then we need to prove that $n^\mu
(\p_\mu\ln\wh\rho-\hbox{str}(\wt\Gamma_\mu))=0$.
To this end, let us introduce
a curve $C$ through $x$ whose tangent vector at $x$ is proportional to
$n^\mu$. Let $t$ parametrize the curve, such that $x(t=0)=x$. We need to
prove that
$$\big(\p_t \ln\wh\rho(x(t)) - \hbox{str}
(\wt\Gamma_t)\big)\big|_{t=0} =0\, .
\eqn\etobeproved $$
The choice  $\rho(x)=\wh\rho(x)$ guarantees that
$F_{x(t)}(x(t)) = \int_{\M_{1,0}} Z_{CFT_{x(t)}}$.
On the other hand, the result of Ref.[\ghoshalsen] together with the fact
that $F_x(x)=\int_{\M_{1,0}} Z_{CFT_x}$, guarantees that $F_x(x(t))=
\int_{\M_{1,0}} Z_{CFT_{x(t)}}$ at least to first order in $t$. This gives,
$$ \p_t \big( F_{x(t)}(x(t)) - F_x(x(t))\big)\big|_{t=0} =0\, . \eqn\efreeone$$
Let us now assume that  $\wh\rho(x)$ does not satisfy
Eqn.\etobeproved. We define,
$$\epsilon(t) = \p_t \ln\wh\rho(x(t))
- \hbox{str} (\wt\Gamma_t) \, . \eqn\efreetwo$$
Let us also define $\wt\rho(t)$ to be a solution of the equation
$$\p_t \ln\wt\rho(t) - \hbox{str} (\wt\Gamma_t) =0\, , \quad
\wt\rho(t=0) =\rho(x) \, ,\eqn\efreethree$$
along the curve $C$. Note that $\wt\rho(t)$ is defined only along the curve
$C(t)$, and hence we are not assuming the integrability of the general set
of equations \ejjjtwo.
In fact, we shall need to assume the existence of
$\wt\rho(t)$ only to first order in $t$.
If we choose $\rho$ to be equal to $\wt\rho(t)$ along the curve $C$, then
by Eqn.\ejjjtwo\ the theory is background independent along this curve.
Let $\wt F_t(t)$ denote the free energy of SFT$_{x(t)}$ around the trivial
background, calculated with this
choice of $\rho$. We then have
$$\p_t \big(\wt F_t(t) - F_x(x(t)\big)\big|_{t=0} =0 \, .\eqn\efreefour $$
(With a slight abuse of notation, we are using $t$ to denote the point
$x(t)$, as arguments of functions that are defined only on $C$.)
Eqs.\efreeone\ and \efreefour\ give,
$$\p_t\big( F_{x(t)} (x(t)) - \wt F_t(t) \big)\big|_{t=0}=0\, .\eqn\efreefive$$
We now note that the computation of $\wt F_t(t)$ and that of $F_{x(t)}(x(t))$
differ from each other only in the choice of $\rho$. Furthermore
${1\over 2}\ln\rho$ appears as an additive term in the expression for $F_x(x)$.
Thus we have
$$ \p_t\big( F_{x(t)} (x(t)) - \wt F_t(t) \big)\big|_{t=0} = \half\,
\p_t \big(\ln \wh\rho(x(t)) - \ln \wt\rho(t)\big) \big|_{t=0}
=\half\, \epsilon(0) \, ,\eqn\efreesix$$
where in the last step we have made use of Eqns.\efreetwo\ and
\efreethree. Comparing Eqns.\efreefive\ and \efreesix\ we see that
$\epsilon(0)$ must vanish. This, together with the definition of
$\epsilon(t)$ given in Eqn.\efreetwo\ proves Eqn.\etobeproved.

\subsection{Finiteness of  $\rho$.}
Let $\{|\Phi_i\rangle\}$ denote a basis for $\H$ comprised at each
ghost number $n$ by sets of basis vectors
$\{|\Phi_r^{(n)}\rangle\}$, and $\{|\wt\Phi_r^{(n)}\rangle\}$,
annihilated by $b_0^+$ and $c_0^+$ respectively,
and satisfying the normalization
condition \ebasisnormalization.
In the subspace of $\H$ defined by the states annihilated by $b_0^+$, the bra
$\bra{G_{12}} \equiv \bra{R'_{12}}(c_0^-c_0^+)^{(2)}$ provides a
canonical, nondegenerate,
symmetric metric  pairing states whose ghost numbers add up to four.
The basis $\{|\Phi_i\rangle\}$ will be said to be of
standard normalization if we further have that
$\bra{G_{12}}\Phi_r^{(n)}\rangle\ket{\Phi_s^{(m)}}=\delta_{rs}\delta_{m+n-4}$.
If we expand the string
field in this basis $\ket{\Psi}=\sum_i\ket{\Phi_i}\psi^i$,
and choose the string field theory measure to be $\rho(x)\prod_i
d\psi^i$, then the propagator in the $b_0^+=0$ gauge is proportional
to $(L_0^+)^{-1}$, and the corresponding
one loop vacuum graph is divergent.

Indeed,  with the conventional kinetic term, the corresponding vacuum graph
is proportional to str$(\ln L_0^+)$ in the subspace of states annihilated
by $b_0$ and $\bar b_0$. This quantity
is manipulated using the relation
$$ \ln(L_0^+) =- \int_0^\infty {d\alpha\over \alpha} e^{-\alpha L_0^+}\,,
\eqn\mnplt$$
which holds up to an infinite constant that is immaterial to the
computation of the supertrace.
We will adjust the basis so that the resulting kinetic term will be
of the form
${1\over g (L_0^+) } L_0^+$ with $g$ chosen so that
$$ \ln \, \Bigl[ {1\over g (L_0^+) } L_0^+ \Bigr] =
 -\int_{2a}^\infty {d\alpha\over\alpha} e^{-\alpha L_0^+}\, .\eqn\fixfun$$
The right hand side of this equation, upon taking the supertrace,
leads to the picture of a vacuum graph built including tori of length
parameter greater than or equal to $2a$ (when the circumference is $2\pi$).
The function $g$ necessary to have \fixfun\ is given by
$$ g(w) = w \, \exp \Bigl( \,\,\int_{2a}^\infty {d\alpha\over\alpha}
e^{-\alpha w}\, \Bigr) \, ,\eqn\requiredg$$
as can be verified by direct substitution.
\foot{ The function $g(w)$
is well defined for $w\ge 0$ (with a finite limit as $w\to
0$), but must be defined by analytic continuation for $w<0$ (as is always
the case when tachyons are present). The analytic continuation can be done
as follows. Let us define
$\phi(w) = \int_{2a}^\infty {d\alpha\over \alpha} e^{-\alpha w}$.
Then ${d\phi(w) \over dw} = -{e^{-2a w}\over w}$,
and we may write $\phi(w) = -\int_\infty^w {e^{-2a z}\over z} dz.$
For $w>0$, the contour of integration can be taken along the real line. For
$w<0$, we take the contour from $\infty$ to $w$ in the upper / lower half
plane, avoiding the origin. The difference between taking the contour in
the upper and lower half plane is given by
$ \int _{C_0} {e^{-2a z}\over z} dz = 2\pi i$,
where $C_0$ is a contour around the origin. This gives rise to an additive
ambiguity of $2\pi i$ in the definition of $\phi(w)$. But $g(w) = w
e^{\phi(w)}$ is well defined. This shows that $g(w)$ has a well defined
analytic continuation to $w<0$.}
We therefore introduce the basis states $\{\ket{\Phi'_i}\}$
related to the properly
normalized basis $\ket{\Phi_i}$ through the relation
$$\ket{\Phi'_i} = (g(L_0^+))^{G_i -5/2} \ket{\Phi_i}\,, \eqn\fedphitr$$
where $G_i$ denotes the ghost number carried by
the state $\ket{\Phi_i}$. The particular form of the scaling was chosen
so that two fields with the same $L_0^+$ eigenvalue and having ghost numbers
that add up to five would scale in precisely opposite ways. Since these are
the states coupled by the symplectic form, such form will remain invariant
under this transformation. Indeed, it is straightforward to verify
that $\omega'_{ij}\equiv (-)\bra{\omega_{12}}\Phi'_i\rangle_{(1)}
\ket{\Phi'_j}_{(2)}= (-)\bra{\omega_{12}}\Phi_i\rangle_{(1)}
\ket{\Phi_j}_{(2)}\equiv \omega_{ij}$.
We now expand the string field as $\ket{\Psi} =\sum_i \ket{\Phi'_i}\psi'^i$
and take the string field theory measure to be of the form
$\rho'(x) \prod_i d\psi'^i$. The kinetic term now takes the form
$$ (-)^{G_i} \bra{R'_{12}} (g(L_0^+))^{-1} (c_0^- Q)^{(2)}
\ket{\Phi_i} \ket{\Phi_j} \,\psi'^i\psi'^j\, , \eqn\ekineticwt$$
and, upon gauge fixing in the $b_0^+=0$ gauge, the kinetic operator
in this primed
basis is given by $(g(L_0^+))^{-1} L_0^+$, as desired. The vacuum graph
calculated from this will be finite (up to tachyon divergence), and will
therefore lead to a finite value for $\rho$ in the primed basis through
the use of Eqn.\erhoexpression.

\chapter{A Batalin Vilkovisky Algebra on Spaces of Riemann Surfaces}

We have introduced in \S2 the antibracket and $\Delta$ operators on
a complex $\wh\P$ spanned by symmetric subspaces of moduli spaces.
In the present section we will enlarge the complex, and introduce
a graded commutative and associative dot product.  A suitably generalized
$\Delta$ operator in this new complex will be seen to give the
structure of a BV algebra. We will be able to give a succint formulation
of the recursion relations for string vertices in terms of the cohomology
of an odd operator $\delta \equiv \p+ \hbar\Delta$ satisfying $\delta^2=0$.
The recursion relations for the $\B$ spaces implementing background
independence can also be written briefly and clearly in terms of a nilpotent
operator $\delta_\V \equiv \delta + \{ \V,\,\}$.

\section{Dot Product}

It is well known that in the space
of functions on a supermanifold, the antibracket
is an object that can be derived if we are provided with
a second order nilpotent derivation $\Delta$ and a graded-commutative
 associative product
($\cdot$) [\wittenab]. Given two functions $A$ and $B$ one has
$$(-)^A \{A,B\} = \Delta (A \cdot B) - (\Delta A)\cdot B
- (-)^A A\cdot (\Delta B) \ .\eqn\abdelta$$
In addition the antibracket acts as a derivation of the dot product, namely
$$\{A, B\cdot C\} = \{A,B\}\cdot C + (-)^{(A+1)B}B\cdot\{A,C\}\,.\eqn\drvcd$$

Since we have found analogs of the antibracket and the delta operator
defined on symmetric subspaces of moduli spaces, one can ask if there
is an analog
for the dot product $(\cdot )$ leading to identities of the above form.
Such object exists, but we have to generalize a bit our definition of
a space of surfaces by including surfaces that can have disconnected
components. Here we shall only sketch the construction of ($\cdot$)
leaving some details for a future publication[\senzwiebachnew].
Let us denote by $\wh\P_{(g_1, n_1)\ldots (g_r, n_r)}$ the
space of surfaces with $r$ disconnected components, with genera $g_1,
\ldots g_r$ respectively, and carrying $n_1, \ldots n_r$ punctures
equipped with local coordinates up to phases. Then $\wh\P_{(g_1,
n_1)\ldots (g_r, n_r)}\equiv \wh\P_{g_1, n_1}\otimes \ldots \otimes
\wh\P_{g_r, n_r}$. We denote by $(\Sigma_1, \ldots \Sigma_r)$ a specific
point in this space where $\Sigma_i\in \wh\P_{g_i, n_i}$. Similarly, if
$\A_i\subset\wh\P_{g_i, n_i}$, we denote by $(\A_1, \ldots \A_r)$ the
subspace $\A_1\otimes\ldots \otimes \A_r$ of $\wh\P_{(g_1, n_1)\ldots
(g_r, n_r)}$. If $\{\A_i\}$ denote the tangent vectors in $\wh\P_{(g_1,
n_1)\ldots (g_r, n_r)}$ induced by the tangent vectors $[\A_i]$ of
$\A_i\subset\wh\P_{g_i, n_i}$, then the orientation of $(\A_1, \ldots
\A_r)$ is taken to be $[\{\A_1\}, \ldots \{\A_r\}]$. {}From this it is clear
that under the exchange of $\A_i$ and $\A_{i+1}$, the orientation of
$(\A_1, \ldots \A_r)$ picks up a factor of $(-)^{\A_i\A_{i+1}}$. Let us
denote by $\wh\C$ the full complex $\oplus \wh\P_{(g_1,
n_1)\ldots (g_r, n_r)}$. Thus $\wh\C$ is the generalization of the complex
$\wh\P$ for surfaces with disconnected components.
The operation of addition has a natural definition  in this complex, and
satisfies $(\A, \B) + (\A, \C) = (\A, \B + \C)$.
{}From the direct product structure of $\wh\P_{(g_1,
n_1)\ldots (g_r, n_r)}$ it is clear that $(\A_1, \ldots \A_n)$ is a zero
element of the complex if any of the $\A_i$ is an empty set.

We now define $\A_1 \circ \A_2 \equiv (\A_1, \A_2)$, and, more generally,
$$(\A_1,\ldots, \A_n)\circ (\A_{n+1}, \ldots \A_{n+m})
\equiv (\A_1,\ldots ,\A_{n+m})\, . \eqn\ebvoneold $$
The $\circ$ operation is bilinear in its argument, and is associative and
graded commutative.
However, note that even if the original elements $\B_1$ and $\B_2$ of the
complex are symmetric in the external punctures, $\B_1\circ \B_2$ is not.
Given two symmetric elements $\B_1, \B_2 \in\wh\C$,
we now introduce the symmetric product:
$$ \B_1 \cdot \B_2 \equiv {\bf S} (\B_1\circ \B_2) \eqn\ebvone $$
where ${\bf S}$ is the same symmetrization operator that appeared in
Eqn.\rbracket. This is the graded commutative associative product in
the complex.

The $\Delta$ and the anti-bracket operations are defined on spaces of
surfaces with disconnected components in a manner identical to the
corresponding operations on spaces of connected surfaces.
In particular, the action of $\Delta$ can result in two
possibilities; the two punctures may lie on the same connected component
of the surface, or
they may lie on two different components of the surface which are
disconnected.
It is then clear that if $\B_1$ and $\B_2$ denote two spaces of surfaces
(containing surfaces with disconnected components in general), then
the object
$$\Delta (\B_1 \cdot \B_2)  - (\Delta \B_1)\cdot \B_2
- (-)^{\B_1} \B_1\cdot (\Delta \B_2) \ ,\eqn\abcelta$$
only contains the surfaces that arise when one surface of $\B_1$
is sewn to another surface in $\B_2$, and therefore is clearly related
to $\{B_1,\B_2\}$, as would be expected from \abdelta. In relating
\abcelta\ to $\{\B_1, \B_2\}$ we need to move the tangent vector
associated with the sewing parameter through the tangent vectors of
$\B_1$, thereby giving rise to the factor of $(-)^{\B_1}$, as is expected
from the left hand side of Eqn.\abdelta.
It is also clear
that for the spaces of surfaces (with disconnected components) the analog
of \drvcd\ holds. The
algebra of the operations $\{ \, ,\,\}\, ,\Delta$ and $(\cdot )$
on $\wh\C$ therefore satisfy the axioms of a BV algebra
[\lianzuckerman,\getzler,\penkavaschwarz].

\section{Representation of the Dot Product On the Space of Functions}

In \S3 we have shown that the $\Delta$ and the $\{,\}$ operations on the
moduli space induce the corresponding opearations in the function space.
It is natural to ask if the same holds for the dot product. To test this,
we first need to extend the definition of $f$ to moduli space of
disconnected Riemann surfaces. Given a subspace
$( \, \A_{g_1,n_1}^{(k_1)}, \ldots \A_{g_r,n_r}^{(k_r)})$ of $\wh\P_{(g_1,
n_1)\ldots (g_r,n_r)}$, we define
$$ f\Big(( \, \A_{g_1,n_1}^{(k_1)}, \ldots \A_{g_r,n_r}^{(k_r)})\Big)
= {1\over \Big(\sum_{i=1}^r n_i\Big)!} \prod_{i=1}^r \Big(
\int_{\A_{g_i,n_i}^{(k_i)}}\bra{\Omega^{(k_i)g_i,n_i}}\Psi\rangle_1\cdots
\ket{\Psi}_{n_i}\Big) \eqn\dotrepone $$
With the help of this equation, and the definition of the symmetrized
product ($\cdot$), it is easy to verify that for any two
elements $\B_1$, $\B_2$ of the complex $\wh\C$, we have
$$ f(\B_1 \cdot \B_2) = f(\B_1) \cdot f(\B_2) \eqn\edotrepone $$
Using this identity, and the corresponding identities involving $\Delta$
and $\{, \}$, we see that the BV algebra of $\Delta$, $\{,\}$ and
$(\cdot)$ on the space of functions is an immediate consequence of the BV
algebra of the corresponding operations in the complex $\wh\C$.

\section{Recursion relations as cohomology conditions}

We shall define the action of the boundary operator $\p$ on a
space of surfaces with disconnected components in a manner identical to
the corresponding operation on the space of connected surfaces. This gives
$$\p (\A_1, \A_2, \cdots \, ,\A_n) \equiv  (\p\A_1, \A_2, \cdots \,,\A_n)+
(-)^{\A_1} \,(\A_1\,,\p \A_2, \cdots \,,\A_n)+\cdots\, , \eqn\newdef$$
and it follows from this definition that $\p$ is a derivation of the
dot product
$$\p\, (\A_1\cdot \A_2) =  (\p\A_1\cdot \A_2)+
(-)^{\A_1} \,(\A_1\cdot\p \A_2)\,. \eqn\newxef$$
We can also define the
exponential function  of an even element $\A\in\wh\C$ by the usual power series
$$\exp(\A) \equiv 1+ \A + {1\over 2} \A\cdot \A + {1\over 3!} \A\cdot\A\cdot\A
+ \cdots \,.\eqn\expdef$$
The object $\exp (\A)$ contains disconnected surfaces even when
$\A$ is a basic space of surfaces. It follows from the last two equations
that
$$\p \,\bigl[ \,\exp (\A)\,\bigr] = \p \A \cdot\exp(\A)\, ,\eqn\diffexp$$
and moreover, using Eqn.\abdelta\ we find
$$ \Delta \exp (\A) = \bigl( \Delta \A + \half
\{ \A , \A\} \bigr) \exp (\A)\,.\eqn\gtmq$$
Let us now introduce the odd operator $\delta$ defined as follows
$$\delta\equiv \p+\hbar\Delta\,.\eqn\newdelta$$
{}From the last three equations we get,
$$ \delta\exp(\A) = \bigl( \p + \hbar \Delta \bigr)
\exp (\A) = \bigl(\p \A + \hbar \Delta \A +
\half\hbar
\{ \A , \A\} \bigr) \exp (\A)\,.\eqn\gtmqx$$

Making use of Eqn.\gtmqx\ we see that we can now write the recursion
relations \recrelnew\ in the simple form
$$ \delta \exp (\V/\hbar) = 0\,.\eqn\sform$$
It is interesting that $\delta$ is actually  nilpotent,
$$\delta^2 = \bigl( \p + \hbar \Delta \bigr)^2 = \p^2 + \hbar (\p\Delta +
\Delta \p)  + \hbar^2 \Delta^2 =0\,.\eqn\newderiv$$
Since $\V$ begins as a
zero dimensional space $\V_{0,3}$, it is fairly clear that $\exp (\V/\hbar)$
is not $\delta$ trivial.
A consistent set of closed string vertices therefore
define a cohomology class of $\delta$
in the vector space spanned by subspaces of moduli spaces of Riemann
surfaces with disconnected components. It can be shown
that the change in $\V$ induced by a change of the cell
decomposition in the moduli space [\hatazwiebach] gives rise to a change in
$exp(\V/\hbar)$ that is $\delta$ trivial [\senzwiebachnew].
Finally, note that if we define the operator:
$$ \delta_{\V} \equiv \delta + \{ \V , \,\cdot \,\}\,, \eqn\nilpotence$$
then
$$ (\delta_{\V})^2 = \Big\{ \p\V +\hbar \Delta\V + \half\{\V, \V\} \, , \,
\cdot \Big\} \,\,.\eqn\nilpotencetwo$$
Thus we see that the recursion relations may also be interpreted as the
criteria of nilpotence of the operator $\delta_{\V}$.

\section{Recursion Relations for $\B$ spaces}

We want to observe that the language developed in the previous subsections
enables us to obtain a better understanding of the recursion relations
that were used to give an inductive construction of the $\B$ spaces.
We had in Eqn.\rwrt\ that
$$\p\B \simeq  \V'_{0,3} + \K\V - \hbar\Delta\B - \{\V,\B\}\,
 -\underline{\V}\,. \eqn\rwrxx$$
Using the definition of $\delta_\V$ in \nilpotence\ we can now rewrite the
above relation as
$$\delta_\V\,\B \simeq  \V'_{0,3} + \K\V -  \underline{\V}\,. \eqn\rrxx$$
This is a simple equation telling us essentially, that $\K\V -\underline\V$
is $\delta_\V$ trivial. The object $\K\V -\underline\V$ represents, for
any fixed genus and number of punctures, the
difference between the string vertex and the object obtained by
adding one more puncture via $\K$ to a vertex having one less puncture.

We verify in a straightforward way the obvious consistency condition
$$\delta_\V^2\,\B \,\simeq \, \delta_\V\V'_{0,3} + \delta_\V\,\K\V -
\delta_\V\underline{\V}\,\simeq 0\, , \eqn\rryx$$
which follows from a little calculation using $\delta_\V\underline{\V}\,= 0$,
and Eqn.\pcomk.

\chapter{Conclusion}

In this paper we have set up the criteria for background independence
of the full quantum closed string field theory, or, in fact,
the criteria for background independence of any quantum theory
formulated in the BV formalism.
We then proved that closed string field theory is a locally background
independent quantum theory. More precisely, we have shown that there
is a symplectic diffeomorphism that maps the appropriate action weighted
measure of the string field theory formulated around a particular conformal
field theory, to the action weighted measure of the string field theory
formulated around a nearby conformal field theory (a theory related to the
original conformal theory via an infinitesimal marginal deformation).

While we have carried out our analysis in the context of closed string
field theory, the case of
Witten's open string field theory can be studied similarly.
If we ignore subtle issues of regularization that seem not to have
been completely resolved [\thornpr ],
quantum background independence
would be established by the symplectic diffeomorphism constructed in \S9
of [\senzwiebach]. This can be seen by noting that i) subtleties aside,
for Witten's open string theory the classical master action coincides
with the quantum master action, ii) there is a symplectic field redefinition
that relates classical master actions obtained from nearby backgrounds
[\senzwiebach], and
iii) this field redefinition is linear in the fields, so that it leaves the
path integral measure $d\mu$ invariant up to a constant multiplicative factor.
This extra multiplicative factor, which may be infinite,
and the subtleties alluded to above, have their origin in
the appearance of closed strings in open string field theory. A physical
way to regulate everything is to work in the context of open-closed
covariant string field theory [\zwiebachoc]. Here the open string
sector is only homotopy associative, and closed string field interactions
must be added in order to satisfy, in a manifestly finite way, the master
equation. It may be of interest to apply our methods to this theory.

Finally, our analysis of background independence
has been restricted to infinitesimal deformations of conformal field
theories. Given two conformal field theories a
finite distance apart, and string field theories formulated around them,
one may be able to integrate the field redefinition for infinitesimal
deformations to find out the finite field redefinition.
As discussed in ref.[\senzwiebach], we may, in principle, encounter
infinities during the process of integration. We must prove the absence of
infinities, and while
there are reasons that suggest this could be the case
[\senzwiebach], a detailed analysis seems important.
This is possibly the most relevant open question in our whole analysis
of background independence. A proper answer may require developing efficient
techniques to deal with spaces of conformal theories.
Of course, on a more fundamental level,
we should focus on how to use the insights obtained proving background
independence in order to construct a manifestly background
independent formulation of string field theory. Perhaps most striking
was the finding of a BV algebra at the level of Riemann surfaces, and
a natural map (a homomorphism) to the BV algebra of string functionals.
This development could open new directions for investigation.

\ack A. Sen would like to thank the Mehta Research Institute, Allahabad,
for hospitality while part of the work was performed.
B. Zwiebach would like to thank H. Sonoda and E. Witten for comments
that helped improve the clarity of the presentation.

\APPENDIX{A}{A: \ Proof of Eqn.\evacuumthreex. }

Consider an arbitrary surface $\Sigma\in\M_{g,0}$. It follows from
\edefomegagzero\ that
$$\eqalign{
\p_\mu\Omega_\Sigma^{g,0} =\,\, &
(-2\pi i) \,D_\mu(\,\wh\Gamma\,)\,
\bra{ \Omega^{(-2)g,1}_{\wh\Sigma}} 0\rangle \cr
=\,\, & (-2\pi i) \Bigl[ \bigl( \,D_\mu(\,\wh\Gamma\,)\,
\bra{ \Omega^{(-2)g,1}_{\wh\Sigma}}\bigr) \ket{0}
+  \bra{ \Omega^{(-2)g,1}_{\wh\Sigma}}\,\bigl(
D_\mu(\,\wh\Gamma\,)\, \ket{0}\bigr) \Bigr]\, .\cr } \eqn\dervacgr$$
The first term in the right hand side can be written as
$$\bigl( \,D_\mu(\,\wh\Gamma\,)\,
\bra{ \Omega^{(-2)g,1}_{\wh\Sigma}}\bigr) \ket{0}= \int_{\K (\wh\Sigma)}
\bra{\Omega^{(-2)g,2}}\wh\O_\mu\rangle\ket{0}\, ,\eqn\diffstt$$
where we have used \edmugomega. In this right hand side the vacuum state
is to be contracted into the original puncture of $\wh\Sigma$, and
$\K(\wh\Sigma)$, as usual, denotes the set of surfaces built by introducing
an extra puncture anywhere outside the unit disk around the only puncture
of $\wh\Sigma$.  In order to simplify this we now argue that in general the
vacuum state deletes punctures not only for surfaces, but also for the
case of forms
$$\bra{\Omega^{(k)g,n+1}_\Sigma}0\rangle =\, (-2\pi i)\,
\bra{\Omega^{(k+2)g,n}_{\pi(\Sigma)}}\eqn\reducep$$
where $\Sigma$ is a genus $g$ surface with $n\hskip-2pt+1$ punctures,
the vacuum state refers to one of the punctures, and $\pi$ denotes the
operation of forgetting that puncture. As an equality of forms it holds
when on the left hand side we act on a set of tangent vectors that deform
$\Sigma$ while on the right hand side we act on the set of tangent vectors
representing the deformations of $\pi(\Sigma)$ induced by the first set
of vectors, as we forget about the chosen puncture. This relation is
proved as follows. The left hand side is given by
$$\bra{\Omega^{(k)g,n+1}_\Sigma}0\rangle
(\wh V_1 ,\cdots \wh V_p )  = N_{g,n+1} \bra{\Sigma\,}{\bf b}({\bf v}_1)
\cdots {\bf b}({\bf v}_p)\,\ket{0}\, ,\eqn\hwtdltp$$
where $N_{g,n}=(-2\pi i)^{3-n-3g}$,
the ${\bf v}$'s are Schiffer vectors, and $p=k+\hbox{dim}(\M_{g,n+1})$.
As reviewed in [\zwiebachlong],
all deformations of the underlying unpunctured Riemann surfaces can be
performed with Schiffer variations around any puncture. We therefore choose
a puncture different from the special one to represent such tangents. All
deformations having to do with moving punctures and changing local coordinates
must use Schiffer vectors based at those punctures. Therefore, we can arrange
so that in \hwtdltp\ any antighost insertion referring to the special puncture
only moves it or deforms its local coordinate. Since the vacuum is annihilated
by $b_{-1}, b_0, b_1 \cdots$ (and the antiholomorphic analogs), all such
antighost insertions dissappear. We
denote the leftover Schiffer vectors by $\pi ({\bf v})$. We
can now push the vacuum state without obstruction all the way into the
surface state $\bra{\Sigma}$ to find
$$\eqalign{
\bra{\Omega^{(k)g,n+1}_\Sigma}0\rangle
(\wh V_1 ,\cdots \wh V_p ) & = N_{g,n+1} \bra{\Sigma\,}0\rangle
{\bf b}(\pi({\bf v}_1))
\cdots {\bf b}(\pi ({\bf v}_p))\,\, ,\cr
&= (-2\pi i)^{-1}\, N_{g,n}\, \bra{\pi(\Sigma )\,}
{\bf b}(\pi({\bf v}_1))
\cdots {\bf b}(\pi ({\bf v}_p))\,\, \cr
&=(-2\pi i)^{-1}\, \bra{\Omega_{\pi(\Sigma )}^{(k+2)g,n}\,}
\,(\, \pi(\wh V_1) ,\cdots \pi(\wh V_p)\, )\,\,, \cr  }\eqn\hwtdltp$$
where we made use of $\bra{\Sigma}0\rangle = \bra{\pi (\Sigma)}$. The
final equality is the equality we wanted to establish (Eqn.\reducep).

Back to Eqn. \diffstt\ we now have
$$\bigl( \,D_\mu(\,\wh\Gamma\,)\,
\bra{ \Omega^{(-2)g,1}_{\wh\Sigma}}\bigr) \ket{0}=\,(-2\pi i)^{-1}
\hskip-6pt\int_{\pi\big(\K(\wh\Sigma)\big)}
\hskip-6pt\bra{\Omega^{(0)g,1}}\wh\O_\mu\rangle\, .\eqn\diffstt$$
The region of integration $\pi\big(\K(\wh\Sigma)\big)$ may be simplified
as follows. We are integrating
over the space of two-punctured surfaces $\K(\wh\Sigma)$, but with the original
puncture (defined in $\wh\Sigma$) deleted. This space is nothing else than
$\K\big(\pi(\wh\Sigma)\big)=\K(\Sigma)$,
minus the space of one-punctured
surfaces $D(\wh\Sigma)$
comprising the set of surfaces $\Sigma$ with one puncture anywhere in the
(would be) unit disk around the original puncture of of $\wh\Sigma$.
This gives
$$\pi\big(\K(\wh\Sigma)\big)=\K(\Sigma) - D(\wh\Sigma) \eqn\epikd $$

Having evaluated the first term on the right hand side of Eqn.\dervacgr\
we now claim that the second term is given by
$$\bra{ \Omega^{(-2)g,1}_{\wh\Sigma}}\,\bigl(
D_\mu(\,\wh\Gamma\,)\, \ket{0}\bigr)\,= \,(-2\pi i)^{-1}\int_{D(\wh\Sigma)}
\bra{\Omega^{(0)g,1}}\wh\O_\mu\rangle\,.\eqn\dervacgrs$$
If so, the last three equations, back in \dervacgr\ give us
$$\p_\mu\Omega_\Sigma^{g,0}= \int_{\K (\Sigma)}
\bra{\Omega^{(0)g,1}}\wh\O_\mu\rangle\,,\eqn\huio$$
and, as a consequence
$$ \p_\mu \int_{\A_{g,0}} \Omega^{(g,0)} = \int_{\K\A_{g,0}} \,
\bra{\Omega^{(0)g,1}} \wh\O_\mu
\rangle \, . \eqn\evacuumthree $$
This is precisely Eqn.\evacuumthreex. It only
remains to justify \dervacgrs. It is simple to see intuitively why this
equation holds. The vacuum state $\ket{0}$ can be written as
$\ket{0}= \bra{0}R\rangle$ where $R$ the symmetric
reflector satisfying $D_\mu(\wh\Gamma)\ket{R}=0$, and where
the bra vacuum $\bra{0}$ can be represented by a one punctured
sphere with local coordinate $z_1(z) =z$ at $z=0$ and with nothing
elsewhere (in particular in $z=\infty$). Therefore, the unit disk around
this coordinate covers `half' the sphere, and in particular, when we
differentiate with the $\wh\Gamma$ connection, we get the integral of
$\O_\mu$ over the other half of the sphere. When we sew back the result
into the Riemann surface $\wh\Sigma$ we get the integral of $\O_\mu$ over
the unit disk around the puncture in $\wh\Sigma$. Quantitatively,
we have
$$\eqalign{
D_\mu (\,\wh\Gamma\,) \ket{0}_{(1)}  &=
\bigl( D_\mu (\,\wh\Gamma\,) \bra{0}_{(2)}\bigr) \ket{R_{12}} =
-{1\over \pi}\int d^2z\,\bra{0_{(2)},z_{(3)}}\O_\mu\rangle_{(3)}\ket{R_{12}}\cr
&= {1\over -2\pi i} \int dx\wedge dy\,\,
\bra{0_{(2)},z_{(3)}}R_{12}\rangle\,\,  b({\p\over \p x})\,
b({\p\over \p y})\ket{\wh\O_\mu}_3\,, \cr}\eqn\proverest$$
where we used the covariant constancy of $\ket{R_{12}}$. The bra
$\bra{0_{(2)},z_{(3)}}$ is the surface state corresponding to a two
punctured sphere (at $0,z$), and the subscripts label state spaces.
In the last step we also made use of the remarks below \edmugomega.
As we sew this into the form in \dervacgrs\ we must bring
$\bra{0_{(2)},z_{(3)}}R_{12}\rangle$ all the way to the surface state.
On the way stand antighost insertions (in the state space $(1)$). Using the
reflector they can be rewritten in the state space $(2)$ and using
$\bra{0_{(2)},z_{(3)}}$ they can be written in the state space $(3)$. After
that they can be pushed to the right. When $\bra{0_{(2)},z_{(3)}}R_{12}\rangle$
hits the surface state $\bra{\wh\Sigma}$ the extra puncture gets deleted
and a new puncture, in the state space $(3)$ appears. In this manner we
obtain (with the correct coefficient) what the right hand side of
\dervacgrs\ stands for. This concludes our proof of \evacuumthreex.

\APPENDIX{B}{B: Proof of Eqn.\epartithree}

In this appendix we shall give a proof of Eqn.\epartithree.
We begin by recalling that $\vev{\O_\mu}^M_{g=1}$ in this
equation needs to be computed in a specific coordinate system $w$,
in which the torus is described by the identification $w\equiv w+1
\equiv w+\tau$.
The surface state $\bra{\Sigma^{(1,1)}}$ corresponding to
a one punctured torus with modular parameter $\tau$ may be defined as
$$ {}_3 \bra{\Sigma^{(1,1)}(\tau, \bar\tau)}\, =\, {}_{123}\bra{V^{
\prime(0,3)}} e^{2\pi i \tau L_0^{(1)}} e^{-2\pi i \bar \tau \bar
L_0^{(1)}} \ket{R_{12}} \eqn\epartiextra $$
The choice of local coordinate at the puncture of $\Sigma^{(1,1)}$ is
induced by the corresponding choice of local coordinate at the special
puncture of $\bra{V^{ \prime(0,3)}}$, but this choice is irrelevant for
our analysis since the state $\ket{\wh\O_\mu}$ will be inserted there.
$\bra{\Omega^{(0)1,1}}$ is constructed from $\bra{\Sigma^{(1,1)}}
\equiv(-2\pi i)\bra{\Omega^{(-2)1,1}}$
using the descent equations
given in refs.[\zwiebachlong,\senzwiebach]:
$$ d \bra{\Omega^{(k-1)1,1}} = (-)^k \bra{\Omega^{(k)1,1}} Q \, ,
\eqn\edescent $$
Using this equation and the definition
$${}_3\bra{\Omega^{(-2)1,1}} = -{1\over 2\pi i}\, \,
{}_3\bra{\Sigma^{(1,1)}(\tau, \bar\tau)} =-{1\over 2\pi i} \,\,
{}_{123}\bra{V^{\prime(0,3)}} e^{2\pi i\tau L_0^{(1)}} e^{-2\pi i\bar\tau
 \overline L_0^{(1)}} \ket{R_{12}} \, , \eqn\edefsigma $$
we get,
$${}_3 \bra{\Omega^{(0)1,1}}= -{1\over 2\pi i} \cdot 4\pi^2 \cdot
d\tau\wedge d\bar\tau  \,\,
{}_{123}\bra{V^{\prime(0,3)}} e^{2\pi i\tau L_0^{(1)}} e^{-2\pi i\bar\tau
 \overline L_0^{(1)}} b_0^{(1)} \bar b_0^{(1)} \ket{R_{12}} \, .\eqn\eooonon$$
We shall now use Eqn.\eooonon\ to calculate $\bra{\Omega^{(0)1,1}}
\wh\O_\mu\rangle$. Although the result is independent of the choice of
coordinate system at the puncture, we shall choose a specific
coordinate system which will allow us to express the result in terms
of the coordinate dependent object $\vev{\O_\mu}^M_{g=1}$.
It turns out that the required choice
of coordinates is the same one introduced in \S2.3. $\V'_{0,3}$
is mapped to an infinite cylinder of circumference $2\pi$ with the symmetric
punctures at the two points at infinity. The local coordinates at those
punctures are fixed by taking as a common coordinate curve an arbitrary
geodesic circle on the cylinder. On that circle we fix a point to be the
special puncture, and define its coordinate curve to be the set of points
at a distance $2\pi$.
Let $w$ denote this coordinate.\foot{Note that with this choice of
normalization
the coordinate disc covers the parts of the torus more than once. In fact,
the curve $|w|=1/2$ just touches itself.}
Then the torus is described in the $w$ coordinate system by the identification
$w\simeq w+1 \simeq w+\tau$. The coordinate $w$ is related to the
local coordinate $z^{(1)}$ around the puncture 1 through the relation
$z^{(1)}=e^{2\pi i w}$. Let us denote the ghost fields in the $w$
coordinate system by $c^w$, $\bar c^w$, and the field $\O_\mu$ in the
$w$ coordinate system by $\O_\mu^w$. The ghost fields $c^w$, $\bar c^w$
now have expansions
$$ c^w(w) ={1\over 2\pi i} \sum_n c^{(1)}_n e^{2\pi i n w}, \quad \quad
\bar c^w(\bar w) =-{1\over 2\pi i} \sum_n \bar c^{(1)}_n e^{-2\pi i n
\bar w} \, . \eqn\cexpansion $$
where, as usual, the superscript $^{(1)}$ denotes that these operators act
on the state space 1.
{}From the above description of $\bra{V^{\prime(3)}}$ we see that
$${}_{123}\bra{V^{\prime(3)}} \wh\O_\mu\rangle_3
=\bra{R_{12}} c^w(w=0) \bar c^w(w=0) \O_\mu^w(w=0) \,.\eqn\evpthom $$
Eqn.\eooonon\ now gives
$$ \bra{\Omega^{(0)1,1}}\wh\O_\mu\rangle = -{1\over 2\pi i} \cdot
d\tau\wedge d\bar\tau  \,\,
\bra{R_{12}} e^{2\pi i\tau L_0^{(1)}} e^{-2\pi i\bar\tau \overline
L_0^{(1)}} b_0^{(1)} \bar b_0^{(1)} \sum_n c^{(1)}_n \sum_m \bar
c^{(1)}_m \O_\mu^w(w=0) \ket{R_{12}} \, . \eqn\efinalmeasure$$
Only the $m=n=0$  term in the sum in the above equation contributes.
The right hand side of Eqn.(C.8) has the form $\bra{R_{12}} A^{(1)}
\ket{R_{12}}$, which, using the definition of $R_{12}$, reduces to the
form $\sum_i (-1)^{\Phi_i} \bra{\Phi^i} A \ket{\Phi_i}$.
We can choose
the basis of states in such a way that either $\ket{\Phi_i}$ is annihilated by
$b_0$ from the left, or $\bra{\Phi^i}$ is annihilated by $b_0$ from the
right. Thus if $A$ contains a $b_0$ without any accompanying $c_0$,
the matrix element $\bra{\Phi^i} A \ket{\Phi_i}$ vanishes identically.
This shows that only the $n=0$ term contributes. A similar argument involving
$\bar b_0$ shows that only the $m=0$ term will contribute.
Using the definition of $Z^G$ and $\vev{\O_\mu}^M_{g=1}$,  we get
$$\int_{\V_{1,1}+\Delta\B_{0,3}'} \bra{\Omega^{(0)1,1}}\wh\O_\mu\rangle
= -{1\over\pi} \int_{\V_{1,0}} d\tau_1\wedge d \tau_2\,\, Z^G
\vev{\O_\mu}^M_{g=1} \,\, . \eqn\epartithreex$$
This is precisely Eqn.\epartithree.

\singlespace
\refout
\end